\documentclass[twocolumn]{aastex62}

\usepackage{amsmath}
\usepackage{graphicx}
\usepackage{lipsum}
\usepackage{hyperref}
\usepackage{mathtools, nccmath}

\newcommand{\dustem}{\texttt{DustEM}}
\newcommand{\rev}{}
\newcommand{\revv}{}

\graphicspath{{./}{figures/}}

\accepted{July 23, 2020}
\submitjournal{ApJ}

\shorttitle{Mid-IR Imaging of WR~112}
\shortauthors{Lau et al.}

\begin{document}

\title{Resolving Decades of Periodic Spirals from the Wolf-Rayet Dust Factory WR~112}

\correspondingauthor{Ryan Lau}
\email{ryanlau@ir.isas.jaxa.jp}

\author{Ryan M. Lau}
\affil{Institute of Space \& Astronautical Science, Japan Aerospace Exploration Agency, 3-1-1 Yoshinodai, Chuo-ku, Sagamihara, Kanagawa 252-5210, Japan}
\author{Matthew J. Hankins}\affil{Division of Physics, Mathematics, and Astronomy, California Institute of Technology, Pasadena, CA 91125, USA}
\author{Yinuo Han}
\affil{Sydney Institute of Astronomy (SIfA), School of Physics, The University of Sydney, NSW 2006, Australia}
\author{Izumi Endo}
\affil{Department of Astronomy, School of Science, University of Tokyo, 7-3-1 Hongo, Bunkyo-ku, Tokyo 113-0033, Japan}
\author{Anthony F.~J.~Moffat}
\affil{Département de physique, Université de Montréal, C.P. 6128, succ. centre-ville, Montréal (Qc) H3C 3J7, Canada; Centre de Recherche en Astrophysique du Québec, Canada}
\author{Michael E.~Ressler}
\affil{Jet Propulsion Laboratory, California Institute of Technology, 4800 Oak Grove Drive, Pasadena, CA 91109, USA}
\author{Itsuki Sakon}
\affil{Department of Astronomy, School of Science, University of Tokyo, 7-3-1 Hongo, Bunkyo-ku, Tokyo 113-0033, Japan}
\author{Joel Sanchez-Bermudez}
\affil{Instituto de Astronomía, Universidad Nacional Autónoma de México, Apdo. Postal 70264, Ciudad de México 04510, Mexico}
\author{Anthony Soulain}
\affil{Sydney Institute of Astronomy (SIfA), School of Physics, The University of Sydney, NSW 2006, Australia}
\author{Ian R. Stevens}
\affil{School of Physics and Astronomy, University of Birmingham, Edgbaston, Birmingham B15 2TT, UK}
\author{Peter G. Tuthill}
\affil{Sydney Institute of Astronomy (SIfA), School of Physics, The University of Sydney, NSW 2006, Australia}
\author{Peredur M. Williams}
\affil{Institute for Astronomy, University of Edinburgh, Royal Observatory, Edinburgh EH9 3HJ, UK}

\begin{abstract}
WR~112 is a dust-forming carbon-rich Wolf-Rayet (WC) binary with a dusty circumstellar nebula that exhibits a complex asymmetric morphology, which traces the orbital motion and dust formation in the colliding winds of the central binary. Unraveling the complicated circumstellar dust emission around WR~112 therefore provides an opportunity to understand the dust formation process in colliding-wind WC binaries. In this work, we present a multi-epoch analysis of the circumstellar dust around WR~112 using seven high spatial resolution (FWHM $\sim0.3-0.4''$) N-band ($\lambda \sim12$ $\mu$m) imaging observations spanning almost 20 years and includes newly obtained images from Subaru/COMICS in Oct 2019. In contrast to previous interpretations of a face-on spiral morphology, we observe clear evidence of proper motion of the circumstellar dust around WR~112 consistent with a nearly edge-on spiral with a $\theta_s=55^\circ$ half-opening angle and a $\sim20$-yr period. 
The revised near edge-on geometry of WR~112 reconciles previous observations of highly variable non-thermal radio emission that was inconsistent with a face-on geometry.
\revv{We} estimate a revised distance to WR~112 of $d = 3.39^{+0.89}_{-0.84}$ kpc based on the observed dust expansion rate and a spectroscopically derived WC terminal wind velocity of $v_\infty= 1230\pm260$ km s$^{-1}$
With the newly derived WR~112 parameters we fit optically-thin dust spectral energy distribution models and determine a dust production rate of \revv{$\dot{M}_d=2.7^{+1.0}_{-1.3}\times10^{-6}$ M$_\odot$ yr$^{-1}$}, which demonstrates that WR~112 is one of the most prolific dust-making WC systems known. 

\end{abstract}


\section{Introduction} \label{sec:intro}

Classical Wolf-Rayet (WR) stars are descendants of massive O-type stars that exhibit fast winds ($\gtrsim1000$ km s$^{-1}$), hot photospheres ($T_*\gtrsim40000$ K) and high luminosities ($L_*\sim10^5$ L$_\odot$; \citealt{Crowther2007}). Despite their extreme environments, a subset of carbon-rich WR (WC) stars have been observed to actively form dust \citep{Gehrz1974,Williams1987}. These dust-forming WC stars, also referred to as ``dustars" \citep{Marchenko2007}, can be prolific sources of dust with production rates ranging from $\dot{M}_d\sim10^{-10}-10^{-6}$ M$_\odot$ yr$^{-1}$ \citep{Zubko1998,Lau2020}. Due to their large dust output and the short evolutionary timescale associated with the onset of the WR phase ($\lesssim$Myr), WC dustars are likely to be early and significant sources of dust \citep{Lau2020}. However, many open questions persist on the nature of dust formation in these systems.

The presence of a binary companion is believed to be a key factor for dust formation in the hostile environment around WC dustars. The strong wind from the WC star interacts and collides with \rev{the} weaker wind from an OB-star companion which creates a dense shock front that cools, forms dust, \rev{and} streams away from the central binary \citep{Williams1990,Usov1991}. Due to the orbital motion of the central binary, newly formed dust propagates radially outward in different directions \rev{corresponding} to the orbital phase of the system. Observational evidence of this phenomena is clearly demonstrated in the changing orientation of the dusty ``pinwheel" revealed in WR~104 by \citet{Tuthill1999}. The morphology and proper motion of the WC dustar nebulae therefore trace the colliding-wind dust formation and orbital configuration of the central binary.

Mid-infrared (IR) imaging observations that are capable of resolving the nebulae around Galactic WC dustars have recently identified gaps in our understanding on their dust formation process. For example, the newly discovered WC dustar known as Apep (2XMM J160050.7–514245; \citealt{Callingham2019}) exhibits inconsistent dynamics between the dust expansion and the WC wind velocities. An even more long-standing mystery is the nature of the apparent broken pinwheel nebulae surrounding the WC dustar WR~112 (CRL 2104; \citealt{Cohen1976,Marchenko2002,Lau2017}). 

\revv{The WR star in WR~112} (as CRL 2104) was initially classified as a WC8 star by \citet{Cohen1976}, who attributed the relative weakness of the 4650-\AA\ C\,{\sc iii-iv} feature to dilution by a luminous, blue companion. However, no stellar absorption lines were observed from the companion. \citet{Massey1983} re-classified WR~112 as a WC9 star based on optical spectroscopy showing the presence of C\,{\sc ii} lines and a strong 5696-\AA\ C\,{\sc iii} feature with a weaker 5802,12-\AA\ C\,{\sc iv} feature. The stellar sub-type is therefore disputed, but we will argue in favor of the initial WC8 classification based the quantitative spectral classification criteria for WC stars by \citet{Crowther1998}.

Although the companion star in WR~112 has not been directly observed, non-thermal radio emission from this system demonstrates that it hosts a colliding-wind binary \citep{Leitherer1997,Chapman1999,Monnier2002}. However, its highly variable non-thermal radio emission and the complex dust morphology of its surrounding nebula complicate our understanding of the dust formation and binary orbital parameters in this system. This issue is compounded by the conflicting interpretations of the dust morphology and \rev{dust} proper motion \citep{Monnier2007,Marchenko2002,Lau2017}.

\rev{In their analysis of a single-epoch N-band ($\lambda\sim12$ $\mu$m) image of WR~112, \citet{Marchenko2002} initially interpreted the WR~112 dust morphology as a near face-on ``pinwheel" spiral similar to WR104 \citep{Tuthill2008}. However, this model is unable to account for several dust features deviating from the face-on morphology. Keck aperture-masking observations by \citet{Monnier2007} probed dust emission features down to $\sim20$ mas and demonstrated that the central region of WR~112 exhibits no obvious spiral structure, but rather a ``horseshoe" structure with two spurs from the central source extending north. These results suggest a morphology more consistent with an edge-on, as opposed to face-on, viewing angle. An edge-on interpretation for WR~112 was initially hypothesized by \citet{Monnier2002} to explain its highly variable non-thermal radio emission. Most recently, \citet{Lau2017} re-interpreted the circumstellar nebula around WR~112 as a consecutive series of ``stagnant" dust shells exhibiting no proper motion. However, \citet{Lau2017} based their interpretation primarily on two N-band observations taken $\sim9$ yr apart where one of the observations was assumed to be rotated by 180$^\circ$. } 

In this work, we present new N-band ($\lambda\sim12$ $\mu$m) imaging data of WR~112 with the Cooled Mid-Infrared Camera and Spectrometer (COMICS) on the Subaru Telescope taken in Oct 2019 and conduct a multi-epoch proper motion and morphological analysis of high spatial resolution N-band imaging observations taken over almost 20 years. We \rev{revise the morphological} interpretation of the spiral geometry of the WR~112 nebula to reconcile the inconsistencies of previous \rev{observational IR } studies.  \rev{With the new results from our geometric spiral model, we revisit the distance estimate to WR~112 and its dust production properties. We also describe how our revised WR~112 geometry can explain the observed non-thermal radio variability. Lastly, we discuss how WR~112's revised orbital and dust production properties compare it to the population of known WC dustars.}

\section{Observations and Archival Data}
\label{sec:2}

\begin{figure}[t!]
    \centerline{\includegraphics[width=0.98\linewidth]{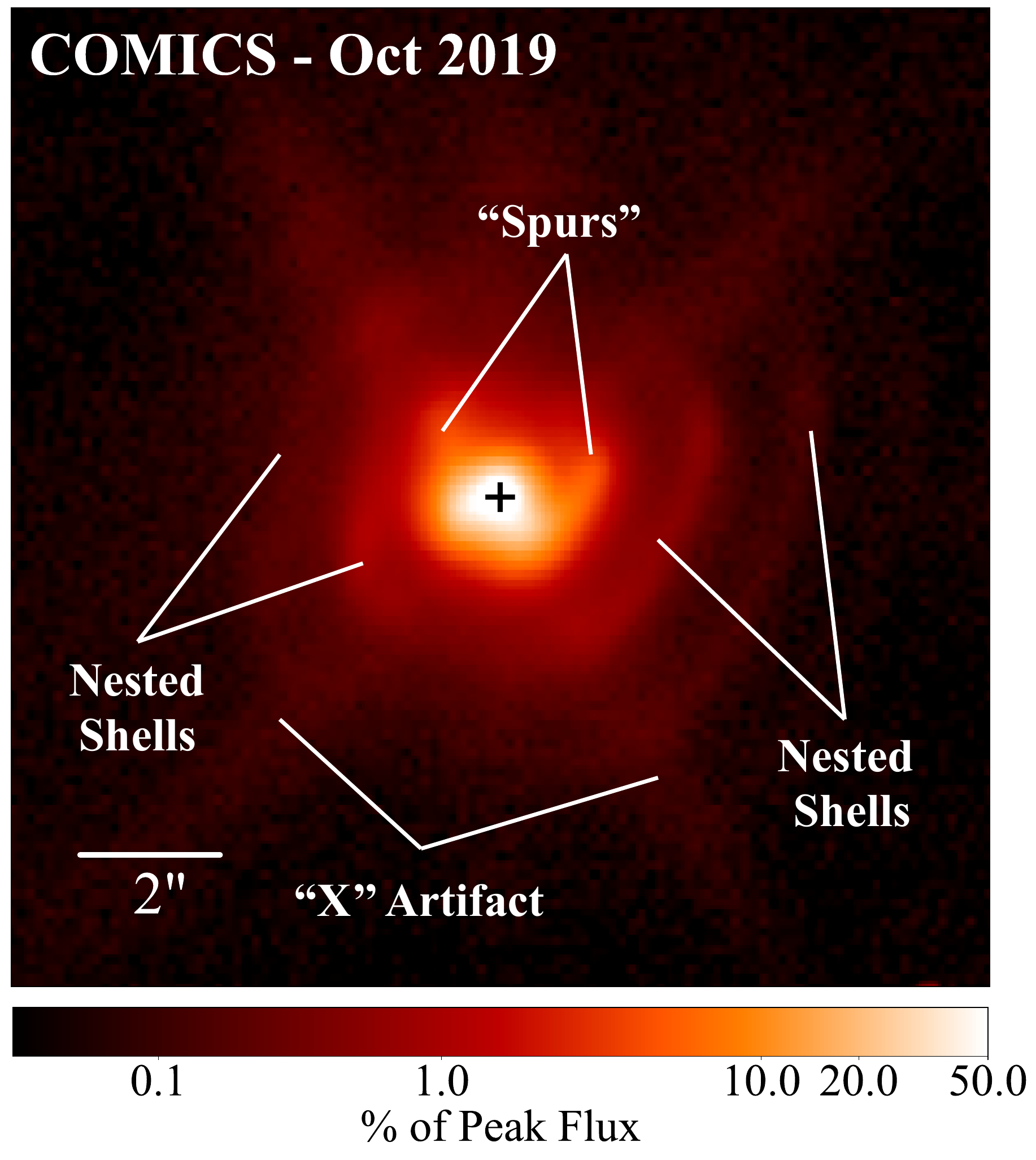}}
    \caption{Logarithmic stretch \rev{from 0.01 - 50\% of peak flux \revv{density}} of the newly obtained Subaru/COMICS N11.7 image of WR~112 overlaid with the features corresponding to the ``spurs" and the nested shell-like emission. The ``X"-shaped artefact due to diffraction by the secondary mirror support struts is also labelled. North is up and east is left.}
    \label{fig:WR112Comics}
\end{figure}

\begin{figure*}[t!]
    \centerline{\includegraphics[width=1\linewidth]{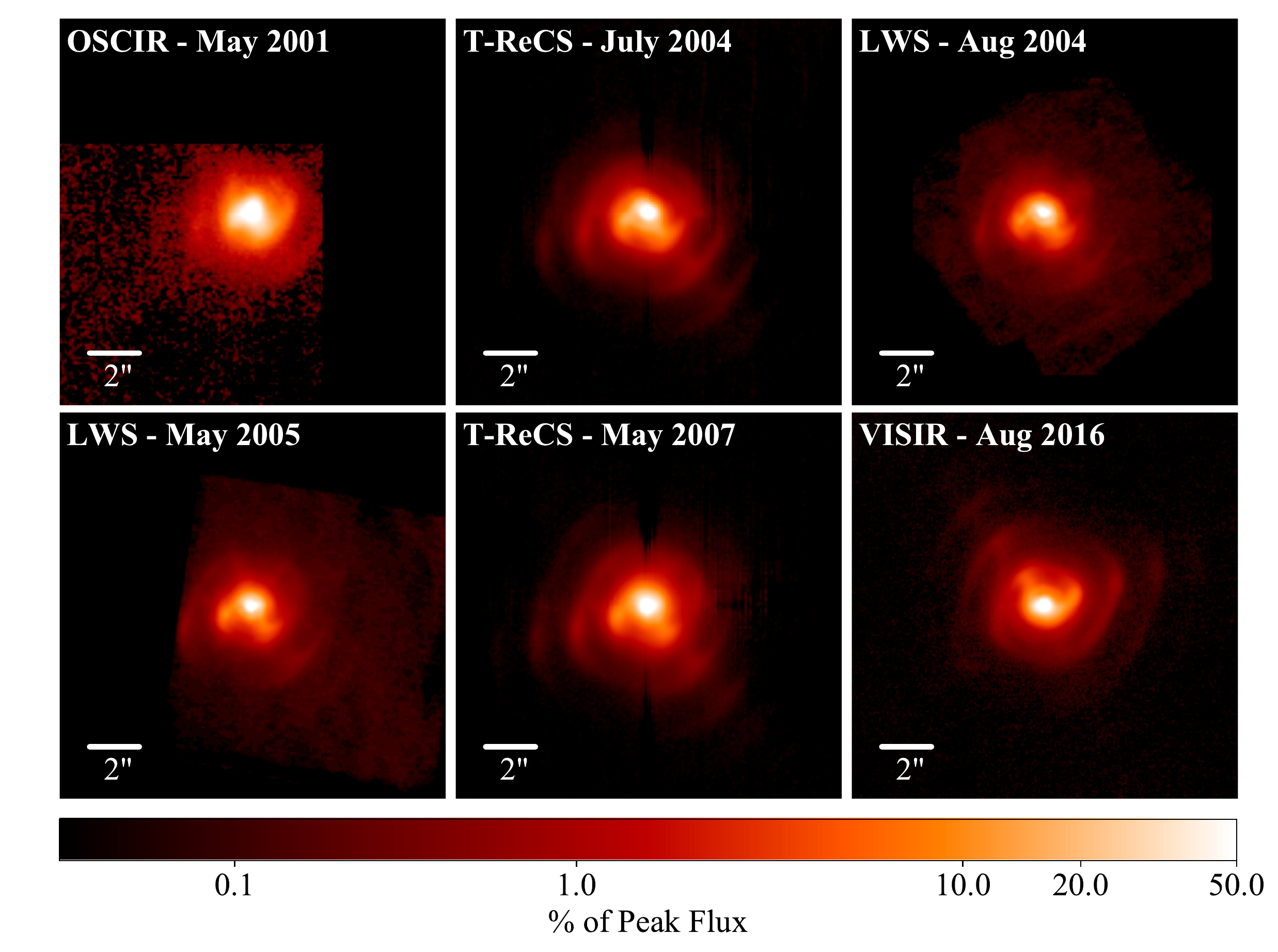}}
    \caption{Logarithmic stretch \rev{from 0.01 - 50\% of peak flux \rev{\revv{density}}} of six previous N-band images of WR~112 taken by Gemini North/OSCIR, Gemini South/T-ReCS, Keck I/LWS, and VLT/VISIR. North is up and east is left in all panels.}
    \label{fig:WR112Im}
\end{figure*}

\subsection{Subaru/COMICS Mid-IR Imaging}

Mid-IR imaging observations of WR~112 (J2000 R.A. 18:16:33.49, Dec. -18:58:42.3; \citealt{Cutri2003}) were obtained using the Cooled Mid-Infrared Camera and Spectrometer (COMICS; \citealt{Kataza2000,Okamoto2003}) on the Cassegrain focus of the Subaru Telescope with the N11.7 filter ($\lambda = 11.7$ $\mu$m, $\Delta\lambda=1.0$ $\mu$m) on 2019 Oct 12 (Fig.~\ref{fig:WR112Comics}). 
Individual 200 s exposures were performed using a perpendicular chopping and nodding pattern with a $20''$ amplitude. The total integration time on WR~112 was 20.1 minutes. The pixel scale of the detector is $0.13''$ pixel$^{-1}$. 

Gamma Aql was used as the photometric and point spread function (PSF) reference from the list of mid-IR standards in
\cite{Cohen1999}.
The measured full-width at half maximum (FWHM) of the calibrator was $\sim0.33''$, which is consistent with near
diffraction-limited performance at 11.7 $\mu$m ($0.29''$).

The data reduction was carried out using COMICS Reduction Software and IRAF. The chopping and nodding pair subtraction was employed to cancel out the background radiation and to reduce a residual pattern \rev{from the} chop subtraction. Flat fielding was achieved using self-sky-flat made from each image. A conversion factor for flux \rev{\revv{density}} calibration was calculated by the reduced images of the standard star and its photometric data \rev{by} \cite{Cohen1999}. The total observed N11.7 flux \rev{density} from WR~112 is $143\pm14$ Jy, where we assume a photometric calibration uncertainty of $\sim10\%$. This is consistent with N-band photometry from previous space- and ground-based observations of WR~112 \citep{Lau2017}.

There are two artefacts present in the COMICS image of WR~112 due to its brightness. The most prominent artefact is the ``X"-shaped emission centered on WR~112 (Fig.~\ref{fig:WR112Comics}), which arises due to diffraction by the secondary mirror support struts (See \citealt{Murakawa2004}). The second, less-prominent artefact is the ``ringing" extending from the center of WR~112 in the row-direction due to cross-talk along the readout channels. The cross-talk artefact is mitigated by the COMICS data reduction routines. Since the morphology of the circumstellar dust emission from WR~112 is distinguishable from the structure of the X-shaped artefact and in order to preserve the signal-to-noise of the extended nebulosity, the analysis on the image is carried out without correction for this artefact.

Hereafter, we refer to the Subaru/COMICS observations as S2019.

\begin{deluxetable*}{llllll}
\tablecaption{\rev{Summary of WR~112 N-band Observations}}
\tablewidth{0.98\linewidth}
\tablehead{Observatory & Instrument & Obs.~Date & Filter Name &  \rev{$\lambda_c$ and $\Delta\lambda$ ($\mu$m)} & \rev{Abbreviation} }
\startdata
Gemini North & OSCIR &  2001 May 7 & 12.5 & 12.49,  \rev{$1.16$} & \rev{G2001}\\
Gemini South & T-ReCS &  2004 July 8 & Si-6 & $12.33$, \rev{$1.18$}& \rev{G2004}\\
Keck I & LWS &  2004 Aug 31 & 10.7 &$10.7$,  \revv{$1.6$}& \rev{K2004}\\
Keck I & LWS &  2005 May 26 & 10.7 &$10.7$,  \revv{$1.6$}& \rev{K2005}\\
Gemini South & T-ReCS &  2007 May 7 & Si-5 &$11.66$, \rev{$1.13$}& \rev{G2007}\\
VLT & VISIR &  2016 Aug 9 & NEII$\_2$ &$13.04$, \revv{$0.20$}& \rev{V2016}\\	
Subaru & COMICS &  2019 Oct 12 & N11.7 &$11.7$, \rev{$1.0$}& \rev{S2019}\\
\enddata
\tablecomments{\rev{$\lambda_c$ and $\Delta\lambda$ correspond to the central wavelength and bandwidth of each filter.}}
\label{tab:WR112Obs}
\end{deluxetable*}

\subsection{Archival Mid-IR Imaging}
\label{Sec:ArchMIR}

Within the past 20 years, WR~112 has been observed with N-band imaging in at least six different epochs in addition to the Subaru/COMICS observations by the following instruments: the Observatory Spectrometer and Camera for the Infrared (OSCIR) at Gemini North on 2001 May 7 (PID - GN-2001A-C-16), the Thermal-Region Camera Spectrograph (T-ReCS; \citealt{trecs}) at Gemini South on 2004 July 8 (PID - GS-2004A-Q-63; PI - A.~Moffat) and 2007 May 7 (PID - GS-2007A-Q-38; PI - J.~Monnier), the Long Wavelength Spectrometer (LWS; \citealt{LWS}) on the Keck I Telescope on 2004 Aug 31 (PID - U18LS; PI - Townes) and on 2005 May 26 (PID - U71LSN; PI - Townes), and the VLT spectrometer and imager for the mid-infrared (VISIR; \citealt{Lagage2004}) on 2016 Aug 9 (PID - 097.D-0707(A); PI - R.~Lau). These six images are shown in Fig.~\ref{fig:WR112Im}, and information on all seven of the N-band observations is summarized in Table.~\ref{tab:WR112Obs}. The filters have different passbands but are sufficiently close in wavelength to probe the same thermal dust emission components from the WR~112 nebula. This is apparent in the identical dust emission morphologies exhibited by WR~112 in 8.6, 13.04, and 19.5 $\mu$m imaging by VISIR \citep{Lau2017}.

Hereafter, we refer to the 2001 Gemini North/OSCIR, 2004 Gemini South/T-ReCS, 2004 Keck I/LWS, 2005 Keck I/LWS, 2007 Gemini South/T-ReCS, and 2016 VLT/VISIR observations as G2001, G2004, K2004, K2005, G2007, and V2016, respectively.

Both K2004 and K2005 images are presented for the first time in this work. \rev{The K2004 and K2005 images were obtained in a non-standard observing mode with varying parallactic angles optimized for an interferometric observing mode, which was the primary focus of the WR~112 program with Keck I/LWS. Standard chopping and dithering were utilized to subtract background emission and mitigate the effects of bad pixels, respectively. The reduced K2004 and K2005 images were obtained by averaging the chop-subtracted, centroid-aligned, and rotation-corrected individual exposures.}

G2001 and G2004 data were published in \citet{Marchenko2002} and \citet{Marchenko2007}, respectively, and G2007 and V2016 data were presented in \citet{Lau2017}. The G2004 and G2007 images were affected by a bright source artefact known as the ``hammer effect"\footnote{See \url{https://www.gemini.edu/sciops/instruments/midir-resources/data-reduction/data-format-and-features}} that causes negative regions extending vertically and horizontally from the bright central core. This effect is mitigated by a median filter subtraction across the image. Although dark vertical bands are still present around the bright central source in the G2004 and G2007 images (Fig.~\ref{fig:WR112Im}), the overall dust emission morphology is readily discernible.

All seven N-band observations, which were obtained on 8 - 10m-class telescopes, are Nyquist sampled and achieve near-diffraction-limited angular resolutions of $\sim0.3 - 0.4''$. OSCIR, T-ReCS, LWS, VISIR, and COMICS use different imager pixel scales of 0.089, 0.09, 0.083\footnote{The official LWS pixel scale is 0.085 arcsec/pixel, but revised calibration of the K2004 and K2005 images indicated a pixel scale closer to 0.083 arcsec/pixel}, 0.045, and 0.13 arcsec/pixel, respectively. The OSCIR, T-ReCS, LWS, and COMICS images were therefore up-sampled to match the 0.045 arcsec/pixel scale of the VISIR image. After re-sampling to the 0.045 arcsec/pixel scale, the images were aligned with the centroid of the bright central core fit by a 2D Gaussian.

\subsection{Archival Mid-IR Spectroscopy}

Archival medium resolution ($R\approx 250 - 600$) 2.2 - 40 $\mu$m spectra of WR~112 were taken by the Short Wavelength Spectrometer (SWS; \citealt{deGraauw1996}) on the Infrared Space Observatory (ISO; \citealt{Kessler1996}) and were obtained in the WRSTARS program (PI - van der Hucht; \citealt{vdh1996}) on 1996 Feb 27. The reduced ``sws" file of the SWS spectrum of WR~112 was obtained from the database of SWS spectra processed and hosted by \citet{Sloan2003}\footnote{\url{https://users.physics.unc.edu/~gcsloan/library/swsatlas/atlas.html}}. The WR~112 spectrum was smoothed by a median filter with a 51-element kernel.

The WR~112 ``pws" file, which shows the spectrum prior to segment-to-segment normalization, did not exhibit significant segment discontinuities greater than $\sim10\%$. However, due to larger flux \rev{\revv{density}} uncertainties beyond 27.5 $\mu$m, only the 2.2 - 27.5 $\mu$m data were used. 

This ISO/SWS spectrum of WR~112 has been previously presented \rev{and/or} analyzed by \citet{vdh1996}, \citet{Chiar2001}, \citet{Chiar2006}, \citet{Lau2017}, and \citet{Marchenko2017}.

\subsection{Optical Spectroscopy and WC8 Spectral Sub-type Confirmation}
\label{Sec:OpSpec}

WR~112 was observed with the Intermediate-dispersion Spectrograph and Imaging System (ISIS) on the William Herschel Telescope (WHT) in the Service Observing program on 1999 July 30 using the R300B grating, which gave a spectral resolution of $\sim5$\AA~for a 1.2" slit. \rev{The observation comprised of five integrations of 1500~s, interspersed with observations of an argon lamp for wavelength calibration. Flat fields were observed at the beginning of the night. The data were reduced using the {\sc figaro} package \citep{Figaro} within the UK Starlink system \citep{Starlink}.}

\begin{figure}[t!]
    \centerline{\includegraphics[width=0.98\linewidth]{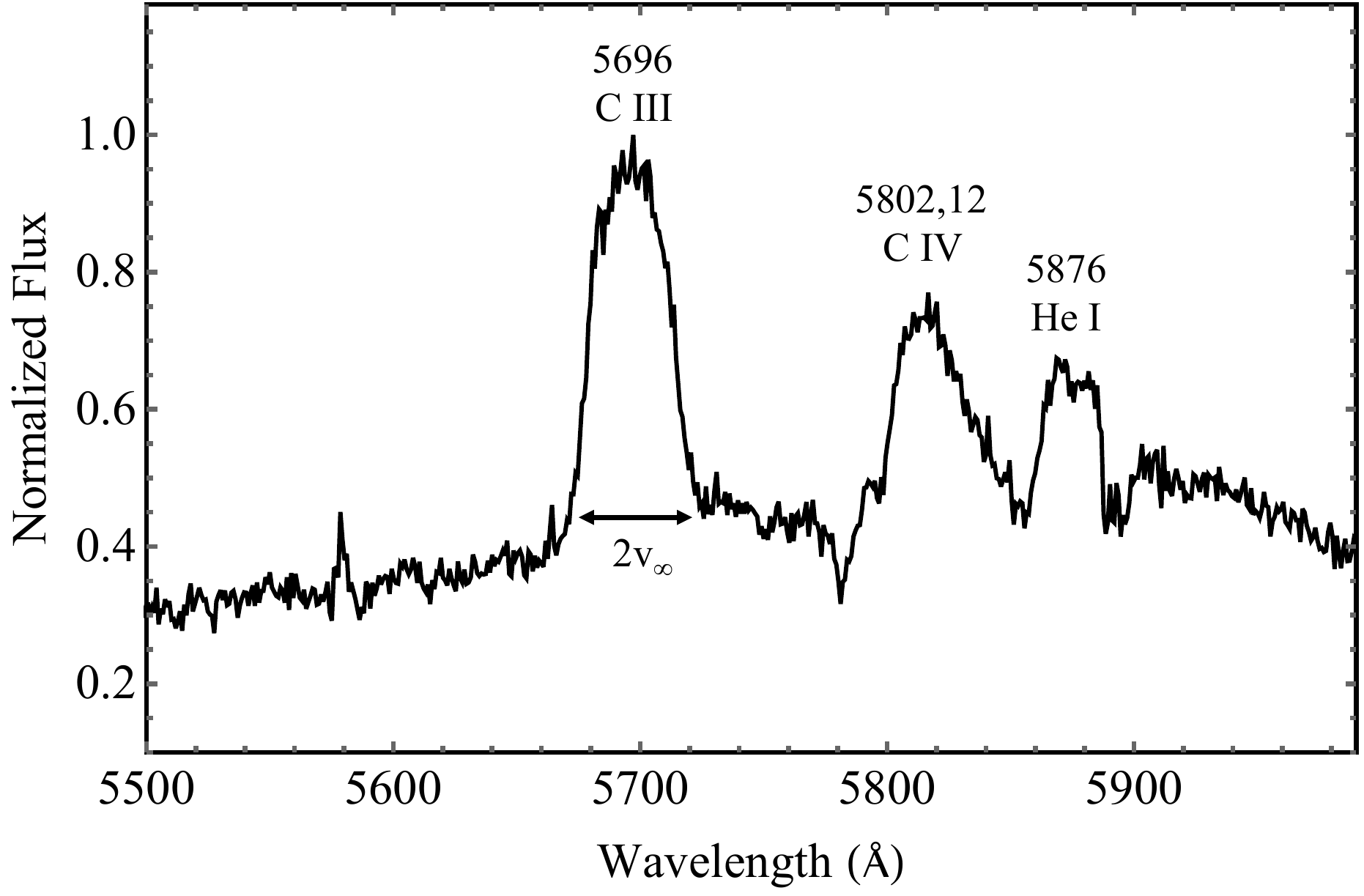}}
    \caption{Optical WHT/ISIS spectrum of WR~112 taken on 1999 July 30 normalized to the peak flux \rev{\revv{density}} of the 5696-$\AA$ C\,{\sc iii} emission line.}
    \label{fig:WR112Spec}
\end{figure}

The spectrum (Fig.~\ref{fig:WR112Spec}) shows prominent carbon and helium emission lines consistent with a late-type WC star. No hydrogen or helium absorption lines indicating an OB companion are present. However, a luminous line-of-sight ``neighbor" is located 942 mas to the SW of WR~112 and is 2.81 mag brighter than WR~112 in the F439W filter taken by the Wide Field and Planetary Camera II on the \textit{Hubble Space Telescope} \citep{Wallace2002}. 
\rev{This neighbor unfortunately falls in the aperture used for the WHT/ISIS spectrum of WR~112 and is also included in the aperture used by earlier spectroscopy by \citet{Cohen1976} and \citet{Massey1983}.} The neighbor dominates the spectrum in the blue, not only diluting the WC emission lines but also any absorption lines from the expected OB companion. 

The prominent 5696-\AA~C\,{\sc iii} line is expected to form in the outer region of the WC wind where it has attained its terminal velocity. The WC wind terminal velocity can then be estimated from the full width at zero intensity (FWZI) of the C\,{\sc iii} line, where FWZI $\approx2v_\infty$. The measured terminal velocity is $v_\infty= 1230\pm260$ km s$^{-1}$.

\rev{The primary quantitative WC sub-type classification criteria from \citet{Crowther1998} is the equivalent width ratio of the 5802,12-\AA\ C\,{\sc iv} and 5696-\AA\ C\,{\sc iii} features. For WC8 stars, \citet{Crowther1998} determine that Log($W_{\lambda}$ (C\,{\sc iv}) / $W_{\lambda}$ (C\,{\sc iii})) $= -0.3$ to $+0.1$. The WHT/ISIS spectrum of WR~112 provide a line ratio of Log($W_{\lambda}$ (C\,{\sc iv}) / $W_{\lambda}$ (C\,{\sc iii})) $= -0.11$, which is consistent with a WC8 classification based on the \citet{Crowther1998} criteria. This corroborates WR~112's initial WC8 classification by \citet{Cohen1976} rather than the re-classification as WC9 by \citet{Massey1983}.}

\section{Results and Analysis}
\label{sec:3}

\subsection{Dust Morphology and Proper Motion}
\label{sec:PM}

The circumstellar dust around WR~112 exhibits a complex morphology with nested, asymmetric shell-like features and two ``spurs" extending from the bright central point source (Fig.~\ref{fig:WR112Comics}). Despite the lack of azimuthal symmetry the successive shells exhibit identical features in the radial direction, which suggests the nested features are related through a periodic formation mechanism.

The WR~112 nebula shows distinct morphological changes over the 18 yr span of the N-band imaging observations. Fig.~\ref{fig:WR112Im} demonstrates the evolving circumstellar dust morphology that is most apparent when comparing the bright, central $\lesssim1.5''$ regions in the G2007 and V2016 images. The central region of the S2019 (Fig.~\ref{fig:WR112Comics}) and V2016 images shows the two spurs extending to the north, whereas the spurs are oriented to the south in the G2004, K2004, K2005 and G2007 images. These central regions show subtle differences between the images taken $\lesssim3$-yr apart. For example, the central morphology in S2019 exhibits a ``W"-like appearance, which slightly deviates from the ``$\nu$"-shaped morphology in the V2016 images. The central region of the G2001 image also exhibits a ``W"-shaped appearance, which was addressed explicitly by \citet{Marchenko2002}.

The G2004, K2004, K2005, G2007, V2016, and S2019 observations exhibit identical $\sim1.5''$ spacings between successive, nested shells located $\gtrsim2''$ from the center. These images reveal emission from at least two dust shells to beyond the two central spurs to the east and west (Fig.~\ref{fig:WR112Comics} \&~\ref{fig:WR112Im}). Only the first nested dust shell was detected in the G2001 image due to the lower SNR compared to the other observations. In the K2004 and K2005 images, some of the outer shells were not observed due to the smaller field of view of and lower SNR of the LWS observations. 

\begin{figure}[t!]
    \centerline{\includegraphics[width=0.98\linewidth]{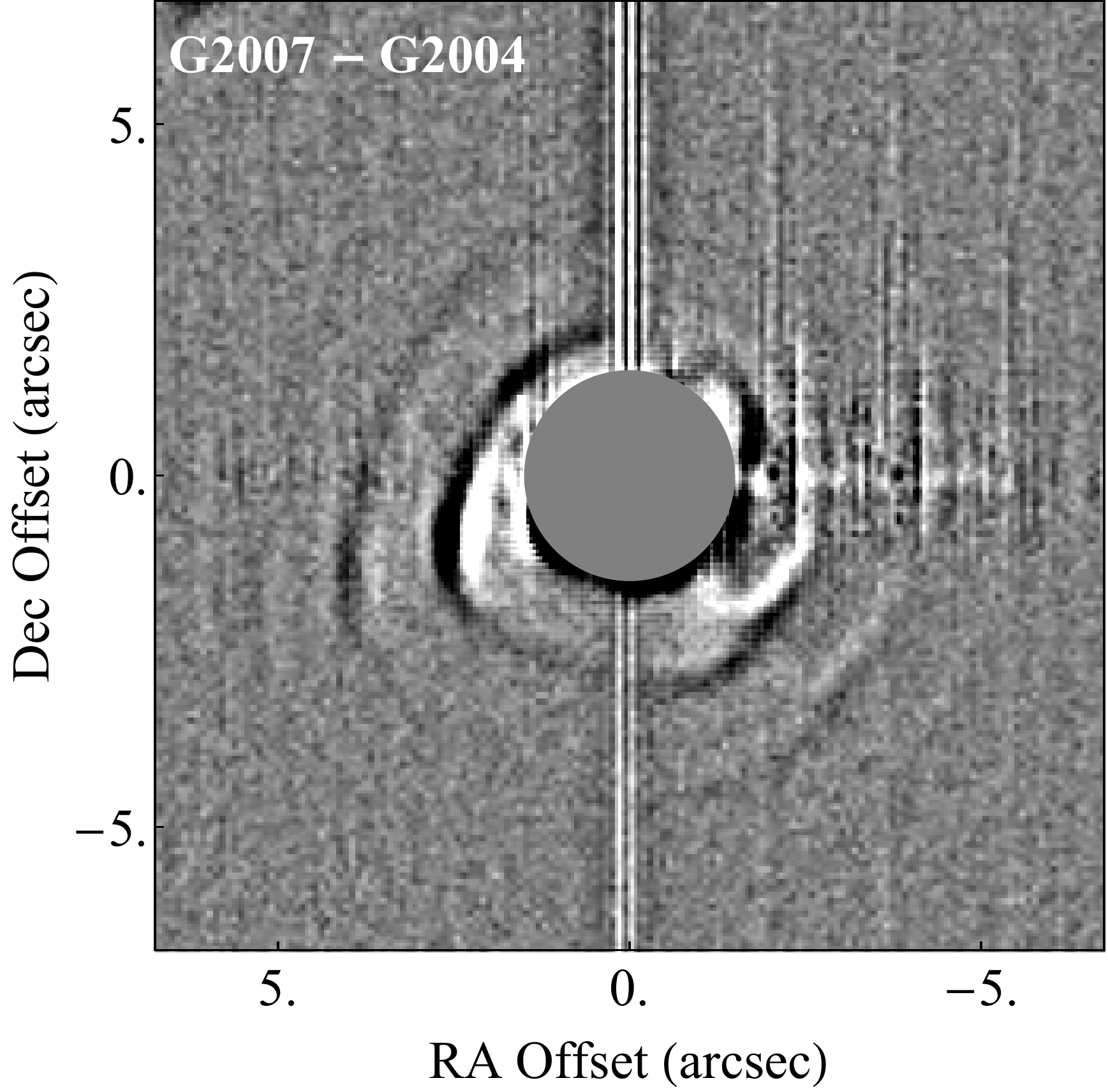}}
    \caption{ \rev{Subtraction residuals of high-pass filtered G2007 and G2004 images. Black and white correspond to positive (G2007) and negative (G2004) emission, respectively. The central region is masked out and the vertical bands extending from the central region are artefacts due to the ``hammer effect" (Sec.~\ref{Sec:ArchMIR}).  }}
    \label{fig:WR112Sub}
\end{figure}

The position of the nested dust shells between the $\lesssim3$-yr observation sequences exhibit slight, but measurable radial proper motion. \rev{For example, Fig.~\ref{fig:WR112Sub} shows the image subtraction residuals between high-pass filtered G2007 and G2004 images.} The \rev{subtraction residuals demonstrate a $\sim0.2-0.3''$} radial proper motion of the eastern and western regions of the first shell beyond the spurs between the G2004 and G2007 images. This is consistent with the positive radial proper motion of the same features between the G2001 and G2004 images as well as the V2016 and S2019 images. 
The dust emission features in the G2004 and K2004 exhibit a nearly identical morphology, which is consistent with the near-contemporaneous timing of the two observations. The WR~112 nebula in the K2005 image also shows a slight radial expansion from the nebula in the previous K2004 image.

\begin{figure}[t!]
    \centerline{\includegraphics[width=0.98\linewidth]{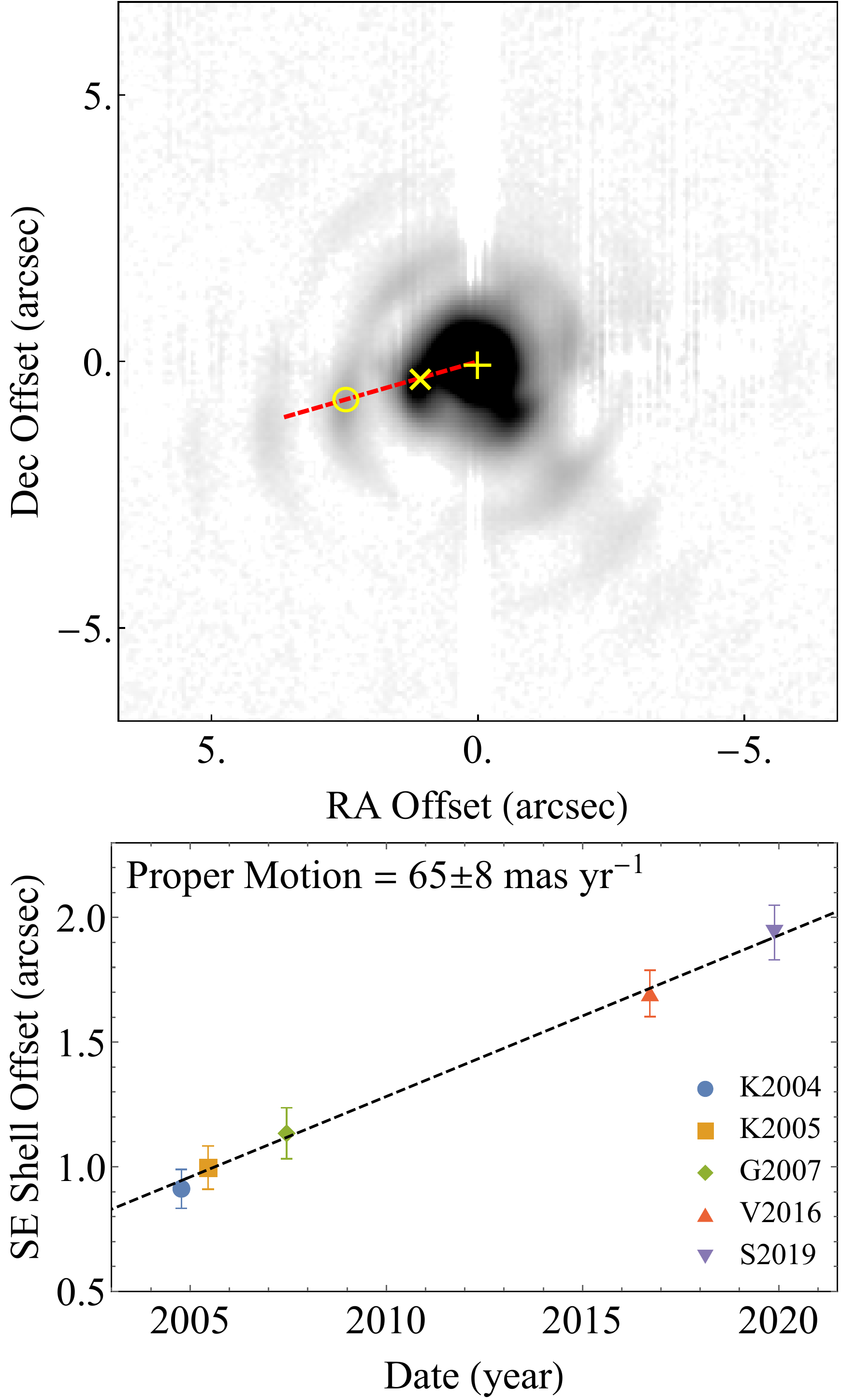}}
    \caption{ (\textit{Top}) Square-root stretch of the median-filter subtracted G2007 image overlaid with the line cut extending from the central core (+) that was used to measure the radial offset of the SE spur/shell ($\times$). The ``$\circ$" marker indicates position of the SE spur/shell in the successive shell. (\textit{Bottom}) Radial offset plot of the SE spur/shell over 5 epochs overlaid with the best-fit line whose slope indicates a proper motion of $65\pm8$ mas yr$^{-1}$.}
    \label{fig:WR112PM}
\end{figure}

The proper motion of the bright southeast (SE) spur from the 2004 - 2007 images can be traced as it expands into the first nested dust shell in the V2016 and S2019 images. The position of the SE spur is indicated \rev{by the ``$\times$" marker in} Fig.~\ref{fig:WR112PM} (\textit{Top}) on a median filter subtracted G2007 image. K2004, K2005, G2007, V2016, and S2019 images were median filter subtracted in order to accentuate the emission from the spur/shell. Radial offsets of the SE spur/shell in the K2004, K2005, G2007, V2016, and S2019 images are measured from the peak location of a Gaussian fit to the emission profile along a radial line cut in the direction of the SE spur (Fig.~\ref{fig:WR112PM}, \textit{Top}). The measured SE spur/shell offsets over 2004 - 2019 are shown in Fig.~\ref{fig:WR112PM} (\textit{Bottom}) and overlaid with a best-fit slope corresponding to a proper motion of $65\pm8$ mas yr$^{-1}$.

Based on the K2004, K2005, and G2007 images, the mean separation distance between the SE spur and the subsequent nested shell (i.e.~the distance between the ``$\times$" and ``$\circ$" markers in Fig.~\ref{fig:WR112PM}, \textit{Top}) is 1.42''. The full ``rotation" period of the repeated shell-like structures can then be estimated from the measured proper motion: $P_\mathrm{PM}=21.8^{+3.1}_{-2.4}$ yr. This $\sim$20~yr period is strengthened by the similar ``W"-shaped appearance of the central regions in the G2001 and S2019 images taken 18.4 yr apart.

\subsection{Geometric Spiral Model}

\rev{\citet{Marchenko2002} initially proposed a face-on spiral morphology for WR~112 with dust expanding radially outward like WR~104 and fit a simple Archimedean spiral with an orbital inclination of $i=38.0\pm3.8^\circ$. However, this face-on model does not appear consistent with the two spurs extending from the central source (Fig.~\ref{fig:WR112Comics}). \citet{Monnier2007} notably identified these spurs in their high angular resolution ($\sim20$ mas) aperture-masking interferometry observations of WR~112. There are additional linear emission features that deviate from a face-on spiral morphology: for example, in the V2016 image (Fig.~\ref{fig:WR112Im}) there are linear features to the northeast that appear to connect the nested shells. These linear features appear in the southwest region of nested shells in the G2004, K2004, K2005, and G2007 images.}

\rev{\citet{Lau2017} claimed that the segmented dust shells had not moved over decades. However, the sequence of images presented in Fig.~\ref{fig:WR112Im} demonstrate that the dust surrounding WR~112 does indeed radial exhibit proper motion, which eliminates the \citet{Lau2017} interpretation of stagnant shells. Here, we revisit the dust morphology of WR~112 and present a revised geometric model.}

We fit a simple 3D conical spiral model to WR~112 defined by the separation between successive spiral turns $\Delta r$, orbital inclination $i$, orbital orientation $\Omega$, \rev{orbital phase} $\varphi$, and half-opening angle $\theta_s$. Given the low orbital eccentricity implied from observations of low-amplitude IR variability from WR~112 \citep{Williams2015,Lau2017}, an eccentricity of $e = 0$ is assumed\rev{; however, geometric models provide an upper limit estimate on the eccentricity of $e\lesssim0.4$. The model we use provides consistent results with the geometric model utilized by \citet{Callingham2019} and \citet{Han2020} in their morphological analysis of the dusty nebula around Apep.}

The basic \rev{($e = 0$)} 3D spiral surface \rev{of our model}, before scaling the size and applying $i$, $\Omega$, and $\varphi$ rotation transformations, is defined by the following set of parametric equations along the $x$, $y$, and $z$ axes:

\begin{equation}
\begin{aligned}
    \rev{x = -u \,\mathrm{Sin}(u) +u\,\mathrm{Cos}(u)\,\mathrm{Cos}(v)\,\mathrm{Tan}(\theta_s)}\\
    \rev{y = -u \,\mathrm{Cos}(u)- u\,\mathrm{Sin}(u)\,\mathrm{Cos}(v)\,\mathrm{Tan}(\theta_s)}\\
    \rev{z = -u \,\mathrm{Sin}(v)\,\mathrm{Tan}(\theta_s)}\\
\end{aligned}
\end{equation}
\noindent
where $0<u<8\pi$ and $0<v<2\pi$. The parameters $u$ and $v$ correspond to the angles along the spiral cone surface that are parallel and perpendicular to the orbital plane, respectively. The $8\pi$ upper range of $u$ therefore traces 4 full ``windings" of the spiral. 
\rev{The orbital phase $\varphi=0$ is defined such that at an inclination $i=90^{\circ}$, our line of sight is aligned with the opening of the ``shock cone" at the apex of the spiral. This $\varphi=0$, $i=90^{\circ}$ spiral orientation corresponds to orbital conjunction where the WC star is aligned directly behind its OB companion along our line of sight (See Sec.~\ref{sec:CW}).}

\begin{deluxetable}{lr}
\tablecaption{WR~112 Geometric Model Results}
\tablewidth{1.0\linewidth}
\tablehead{Parameter & \colhead{\hspace{5.0cm}Value } }
\startdata
$P_\mathrm{PM}$ & $21.8^{+3.1}_{-2.4}$ yr\\
$P_\mathrm{Mod}$ & $19.4^{+2.7}_{-2.1}$ yr\\ 
$\Delta r$ & $1.48\pm0.10''$\\ 
$\theta_s$ & $55\pm5^\circ$\\
$i$  & $100\pm15$$^\circ$ \\
$\Omega$ & $75\pm10$$^\circ$ \\
\rev{$\varphi_\mathrm{G2001}$} & \rev{$0.25\pm 0.14$}\\ 
\rev{$\varphi_\mathrm{G2004}$} & \rev{$0.47\pm 0.11$}\\
\rev{$\varphi_\mathrm{K2004}$} & \rev{$0.47\pm 0.11$}\\
\rev{$\varphi_\mathrm{K2005}$} & \rev{$0.50\pm 0.11$}\\
\rev{$\varphi_\mathrm{G2007}$} & \rev{$0.58\pm 0.11$}\\
\rev{$\varphi_\mathrm{V2016}$} & \rev{$0.08\pm 0.08$}\\
\rev{$\varphi_\mathrm{S2019}$} & \rev{$0.22\pm 0.11$}\\
\enddata
\tablecomments{Summary of the geometric spiral properties of WR~112, where a circular ($e=0$) orbit was assumed. $P_\mathrm{PM}$ and $P_\mathrm{Mod}$ are the orbital periods derived from the dust proper motion analysis and the geometric model, respectively. $\Delta r$, $\theta_s$, $i$, and $\Omega$ are the separation distances between the successive spiral turns, the half-opening angle of conical spiral, the orbital inclination, and the orbital orientation, respectively. $\varphi$ shows the \rev{orbital phase} fit to the seven different epochs.}
\label{tab:WR112OrbResults}
\end{deluxetable}

\begin{figure*}[t!]
    \centerline{\includegraphics[width=1\linewidth]{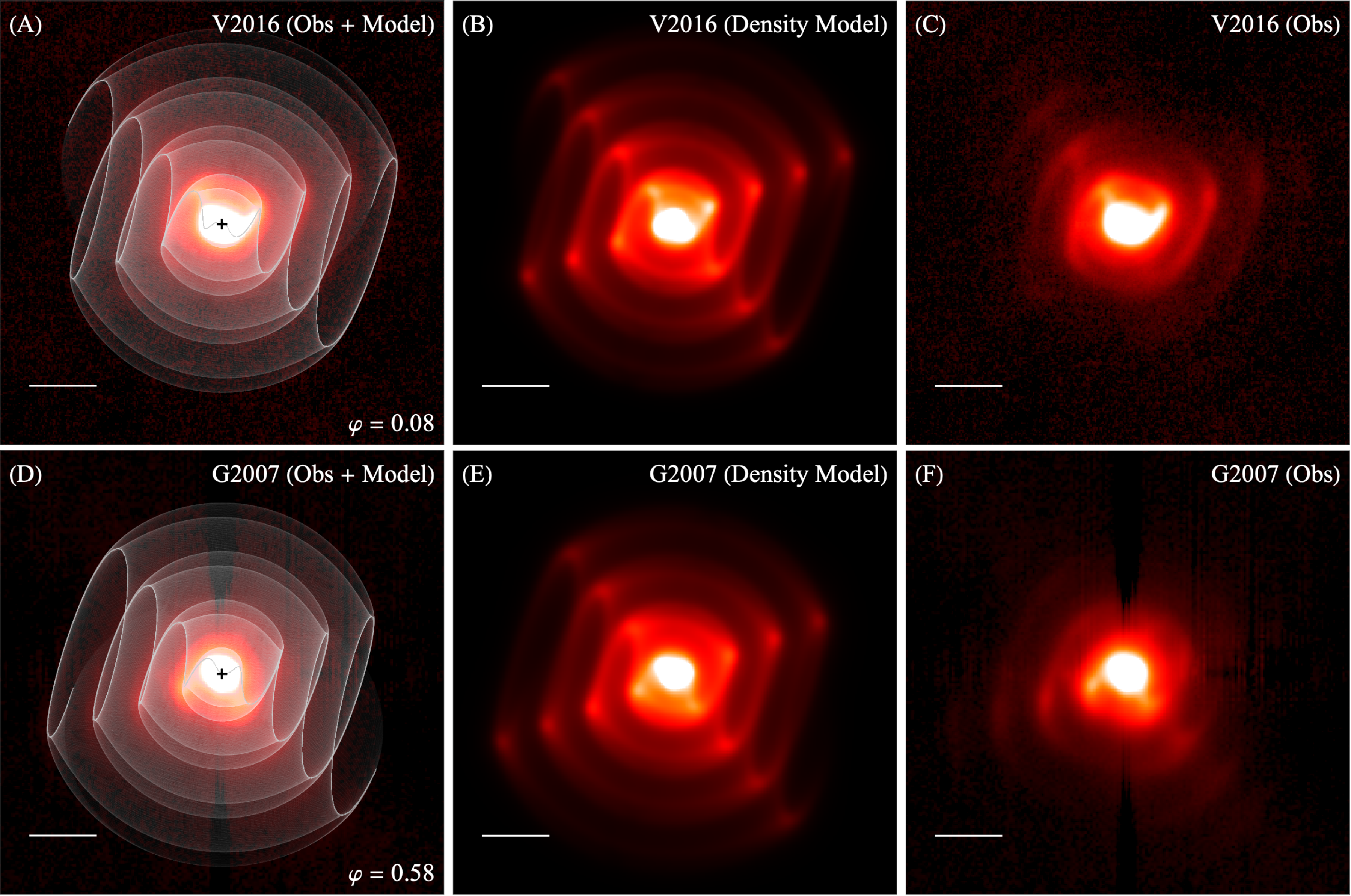}}
    \caption{(A) Square-root stretch of the V2016 image of WR~112 overlaid with the V2016 geometric spiral model \rev{($\varphi = 0.08$)}. The black cross corresponds to the center of the model and is aligned with the emission centroid of the central region. (B) Linear stretch of the dust column density model image at the V2016 epoch. (C)  Square-root stretch of the V2016 image. Panels (D), (E), and (F), correspond to the same descriptions as (A), (B), and (C) for the G2007 observation and model \rev{($\varphi = 0.58$)}. In all panels, north is up and east is left, and the length of the overlaid white lines correspond to 2''. \rev{An animated version of this figure is available that shows (Left) the evolution of the density model image (convolved with the VLT/VISIR PSF) over an orbital period from $\varphi = 0$ to 1 with intervals of $\Delta\varphi\approx0.028$ and (Right) a corresponding sequence of V2016, S2019, K2004, and G2007 images with fading transitions that pause for 2 seconds at the matching model phase. In the both animation panels north is up and east is left, and the length of the white lines correspond to 2''. The total duration of the animation is 13 seconds.}}
    \label{fig:WR112Mod}
\end{figure*}

These parameters are initially fit to the dust emission from the V2016 image, which provides the highest signal-to-noise ratio detection of the extended dust features with minimal imaging artefacts. The models were fit by aligning the \rev{2D projection of the} edges of the 3D spiral to the dust emission features in the images (e.g.~Fig.~\ref{fig:WR112Mod}A). The G2001, G2004, K2004, K2005, G2007, and S2019 images are then fit by adjusting only the \rev{orbital phase} $\varphi$ with the fixed V2016 parameters for $\Delta r$, $i$, $\Omega$, and $\theta_s$. The $\varphi$ parameters and the dates of the seven observations were used to derive the rotation period of the spiral.

Our fitted model presents the WR~112 nebula as a near edge-on spiral ($i=100^\circ$) with a wide half-opening angle of $\theta_s=55^\circ$ where the dust emission corresponds to the edges of the spiral plume. \rev{This revised geometry is notably consistent with the edge-on hypothesis suggested by \citet{Monnier2002,Monnier2007}.} Four full spiral turns of the V2016 spiral model is shown in Fig.~\ref{fig:WR112Mod} (A) overlaid on the V2016 image, and the spiral model parameters are provided in Tab.~\ref{tab:WR112OrbResults}. Uncertainties in the model parameters correspond to the upper and lower ranges where the projected spiral edges no longer align with the dust emission from the nebula. 

The model fits show that the two spurs extending from the central source are associated with the edges of the spiral plume (Fig.~\ref{fig:WR112Mod}A). Notably, the extended linear emission that connects the nested shells in the V2016 image is reproduced by the projection of the eastern edge of the successive spiral turns. The deviations from circular, azimuthal symmetry are also explained by the high inclination. The model fit to the G2007 image (Fig.~\ref{fig:WR112Mod}D) also reproduces to the observed asymmetric features at a nearly opposite orientation from the V2016 image.

The rotation period of the spiral is $P_\mathrm{mod}=19.4^{+2.7}_{-2.1}$ yr based on a least-squares fit to the seven $\varphi$ values of each observation with respect to the observation dates. This is consistent with the $\sim20$ yr period estimate based on the measured proper motion and the $\sim1.4''$ intervals between the dust arcs (Sec.~\ref{sec:PM}). Since there is a direct link between the orbital phase of the central dust-forming binary system and the position angle of the dust plume \citep{Monnier1999,Tuthill2008}, we infer a 19.4 yr orbital period equal to the rotation period for the central binary system of WR~112.

\subsection{Model Dust Column Density Image}

We produce a 2D dust column density image using the geometry of the fitted 3D spiral in order to compare to the observed dust morphology of the WR~112 nebula. Dust is assumed to be confined to the geometrically thin surface of the 3D conical spiral where the dust density decreases as a function of radius from the central system $n_d\propto r^{-2}$.
The dust density image is derived from a 2D projection of the 3D geometric spiral, where dust in the spiral is summed along the line of sight in columns consistent with the VISIR pixel scale of $0.045\times0.045''$. 

The V2016 density model image is presented in Fig.~\ref{fig:WR112Mod} (B), which has also been convolved with the VLT/VISIR NeII$\_2$ point spread function (PSF) determined from selected standard stars observed in 2016 \citep{Lau2017}. Four full spiral turns are modelled in the density image, which is consistent with the maximum number of observed spiral turns. The surface brightness of the fourth spiral in the V2016 image is less than $1\%$ of the central region; therefore, the inclusion of additional spiral turns should not significantly alter the density image. 
\rev{It is important to note that the density image effectively models dust emission under the assumption of constant dust temperature and does not reproduce the exact observed dust emission profile from the WR~112 nebula, where dust temperature decreases as a function of radius from the central system ($T_d\propto r^{-0.4}$; \citealt{Marchenko2002,Lau2017}). However, it is still informative to compare the morphology and column density projection effects from the model image to the observed dust emission.}

The morphology and limb brightening in the V2016 density model closely reproduces the dust emission in the V2016 image (Fig.~\ref{fig:WR112Mod}B and C).
The observed nested shell-like emission beyond the central spurs indeed correspond to the limb brightened edges of the 3D spiral model. Notably, the locations of high column density ``corners" in the density image where the spiral curves and aligns with our line of sight is consistent with regions of locally enhanced dust emission in the V2016 image. 

A comparison of the G2007 density model and image (Fig.~\ref{fig:WR112Mod}E and F) \rev{shows} the same close agreement. The timing of the G2007 observations notably corresponds to a nearly 180$^\circ$ phase shift from the V2016 observations. The rotated appearance of the spurs and dust emission asymmetries are reproduced by the G2007 density model. The striking resemblance of the density models to the images reinforces our revised interpretation of the WR~112 nebula.

\rev{One notable discrepancy between the density model and the observations is that dust emission to the north and south of the central system appears weaker in the observed images. This effect is most noticeable in the northern regions of the outer shells in the V2016 image and the southern regions of the outer shells in the G2007 image (Fig.~\ref{fig:WR112Mod}C and F). We note that the ``isothermal" density model does not reproduce the effects of radiative heating, which could be responsible for the weaker north and south dust emission. The radiation field from the central binary may indeed be asymmetric given that the dimmed regions of the nebula are aligned with the orbital plane where the impinging radiation may be attenuated by the dense wind-collision region between the WC star and its OB companion.}

\begin{figure}[t!]
   \centerline{\includegraphics[width=0.98\linewidth]{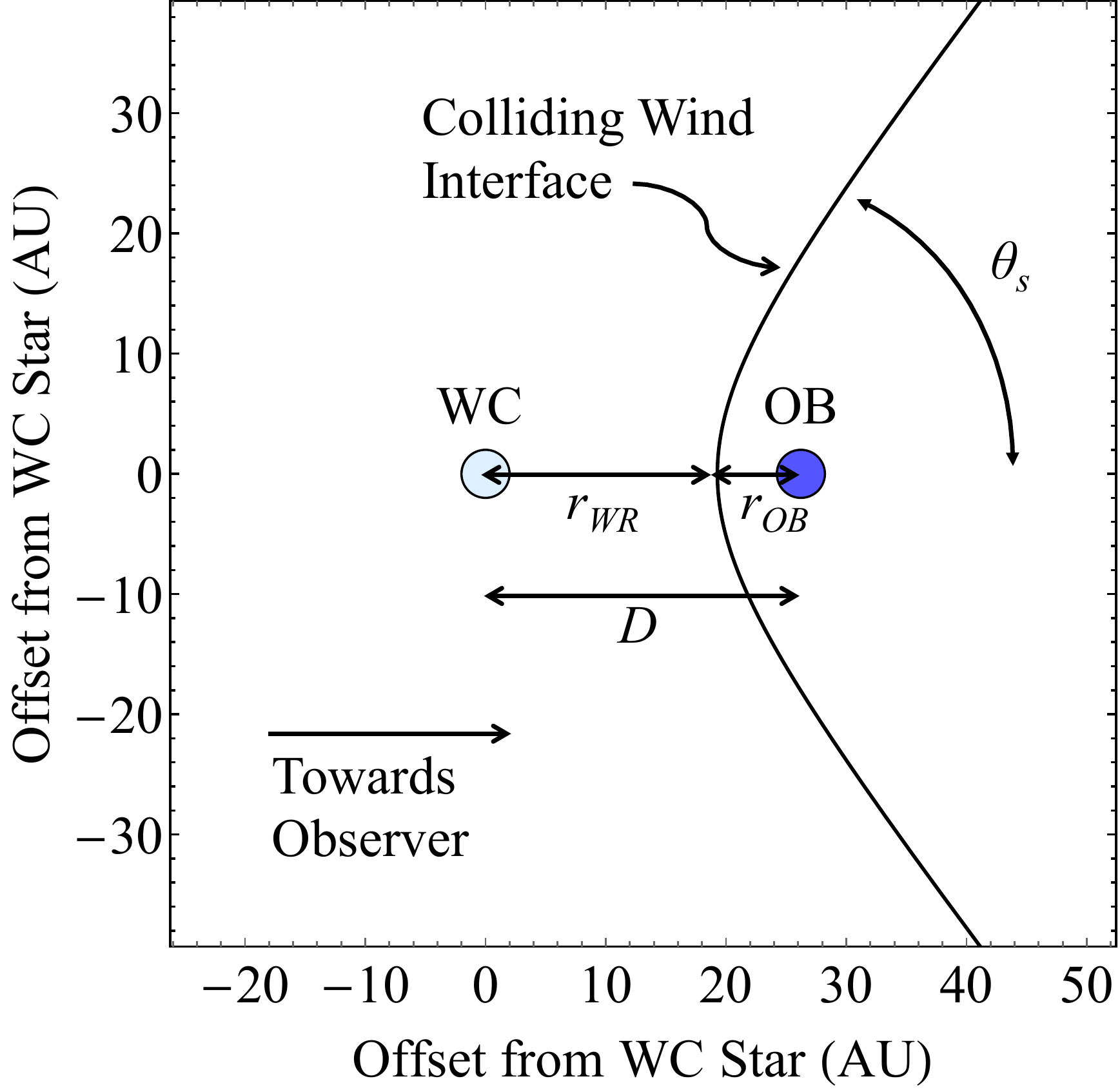}}
    \caption{Face-on diagram of the central WR~112 WC+OB binary and the colliding wind interface with the origin centered on the WC star. The size of the WC and OB star circles are not shown to scale. The orbital separation, $D$, separation between the WC/OB star and the apex of the shock cone, $r_{WR/OB}$, and the half-opening angle of the shock cone, $\theta_s$, are all shown to scale. }
    \label{fig:CWSchem}
\end{figure}

\subsection{The Central Binary and Wind Collision Interface}
\label{sec:CW}
Our revised geometric model of the WR~112 nebula allows us to investigate the configuration of the central WC+OB binary and the shock cone formed by their colliding winds. In Fig.~\ref{fig:CWSchem}, we present a face-on projection of the central binary and the colliding-wind interface of WR~112. Due to the dominant momentum of the WC-star wind over the \rev{OB}-star wind, the ``shock cone" opens in the direction of the OB star.

The geometry of the wind interface in Fig.~\ref{fig:CWSchem} is described by the two-wind interaction model by \citet{Canto1996}, which assumes a purely hydrodynamical balance between the winds. In this model, the half-opening angle $\theta_s$ of the colliding-wind shock cone can be used to derive the wind momentum ratio $\eta \equiv \frac{\dot{M}_{OB} v_{OB}}{\dot{M}_{WR} v_{WR}}$ as follows \citep{Canto1996}:

\begin{equation}
\label{eq:eta}
    \theta_\infty - \mathrm{tan}\,\theta_\infty = \frac{\pi}{1 - \eta},
\end{equation}
\noindent
where $\theta_\infty=\pi-\theta_s$.
Given an opening angle of $\theta_s=55^\circ$, the wind momentum ratio in WR~112 is $\eta \approx 0.13$. The wind momentum ratio in WR~112 is notably \rev{higher} than the ratio in face-on pinwheel system WR~104 ($\eta \approx 0.003$; \citealt{Tuthill2008}), which also exhibits a narrower half-opening angle ($\theta_s\approx20^\circ$) than WR~112.

The orbital separation and the distances between the WC and OB star and the wind collision region can be estimated from the 19.4 yr orbital period and the wind momentum ratio. Assuming a circular orbit, \rev{a WC star mass of 18 M$_\odot$, which is consistent with the mean Galactic WC8 stellar mass derived by \citet{Sander2019}, and an OB companion mass of $30$ M$_\odot$,} the orbital separation of the WR~112 central binary is \rev{$D\approx26$ AU}. The distances $r_{WR}$ and $r_{OB}$ from the WR and OB star to the wind collision region can then be derived from the following relation by \citet{Usov1991}:

\begin{equation}
\label{eq:Rat}
    r_{WR}=D\frac{1}{1+\eta^{1/2}}\,\mathrm{,}\,\,
    r_{OB}=D\frac{\eta^{1/2}}{1+\eta^{1/2}}.
\end{equation}
\noindent
It follows from Eq.~\ref{eq:Rat} that the separation distances between the WR and OB star from the wind collision front are \rev{$r_{WR}\approx19$ AU} and \rev{$r_{OB}\approx 7$ AU}. The separation distances as well as the colliding-wind interface are shown to scale in Fig.~\ref{fig:CWSchem}.

Lastly, it is important to note that the orbital motion of the central binary will lead to ``aberrations" of the apex of the shock interface \citep{Parkin2008}, which is not incorporated in the shock cone geometry shown in Fig.~\ref{fig:CWSchem} since its effects in WR~112 are negligible. The effect of these aberrations can be characterized by a skew angle $\mu$ between the binary axis and the symmetric axis of the colliding-wind interface: $\mathrm{Tan}(\mu)=(v_\mathrm{orb}/v_\infty)$, where  $v_\mathrm{orb}$ is the orbital speed of the binary \citep{Parkin2008}. For the central binary of WR~112, \rev{$v_\mathrm{orb}\approx40$ km s$^{-1}$} and $v_\infty\approx1230$ km s$^{-1}$, which demonstrates that the skew angle \rev{$\mu\approx1.9^\circ$} is indeed small.


\subsection{\rev{Revised WR~112 Distance Estimate}}
\label{Sec:dist}

\rev{Previous} distance estimates towards WR~112 \rev{are highly uncertain and} range from $1.3 - 4.15$ kpc \citep{Nugis1998,vdh2001} due to the large interstellar extinction ($A_v = 12.24$; \citealt{vdh2001}), its dusty nebula, and luminous neighbor \citep{Wallace2002}. 
A recent study by \citet{RC2020}, who perform a Bayesian analysis on Gaia DR2 parallaxes to 383 Galactic WR stars with priors based on H\,II regions and dust extinction, obtains a distance of $d=3.16^{+2.06}_{-1.07}$ kpc. However, \citet{RC2020} flag WR~112 for exhibiting a large negative parallax and high astrometric excess noise $>1$ mas.

\rev{Similar to previous studies of WR~104 \citep{Tuthill2008,Soulain2018}, we can estimate the distance to WR~112 by assuming the observed dust expansion velocity is consistent with the terminal velocity of the WC wind. Based on the model period $P_\mathrm{Mod}$ and the intervals between the spiral turns from the geometric model $\Delta r$, the model-derived dust expansion rate is $\Delta r/ P_\mathrm{Mod} = 76^{+12}_{-10}$ mas yr$^{-1}$. Given the $v_\infty=1230\pm260$ km s$^{-1}$ terminal wind velocity measured from the FWZI of the 5696-$\AA$ C\,{\sc iii} line in the WHT/ISIS spectrum (Fig.~\ref{fig:WR112Spec}), we derive a distance to WR~112 of \revv{$d=3.39^{+0.89}_{-0.84}$ kpc.}}

\rev{This method of deriving WC dustar distances is notably challenged by the discovery of discrepant WC wind and dust expansion velocities in the WC dustar Apep, where the spectroscopically measured line-of-sight WR wind velocity is a factor of $\sim4$ higher than the observed dust expansion velocity \citep{Callingham2019,Callingham2020,Han2020}. Apep's velocity discrepancy is attributed to a rapidly rotating WR star that exhibits a slow equatorial wind, where dust forms via wind-collision with its companion, and a fast polar wind, where dust formation is absent \citep{Callingham2019,Callingham2020}. Assuming spin-orbit alignment, the spectroscopically measured fast wind is indeed consistent with our line-of-sight alignment towards Apep's pole given the near face-on inclination of its orbit ($i\approx25^{\circ}$; \citealt{Han2020}). Apep's viewing geometry notably differs from WR~112's, which we interpret as near edge-on. Therefore, even if the WC star in WR~112 exhibits a discrepant equatorial and polar wind, its spectrum would be dominated by the same equatorial wind along our line-of-sight that condenses into dust via colliding-winds from its companion. We therefore claim that the distance estimate method of equating the measured WC wind terminal velocity to the observed dust expansion velocity is valid for WR~112.}

\subsection{Revisiting Dust Production Properties from WR~112}

\begin{figure}[t!]
    \centerline{\includegraphics[width=0.98\linewidth]{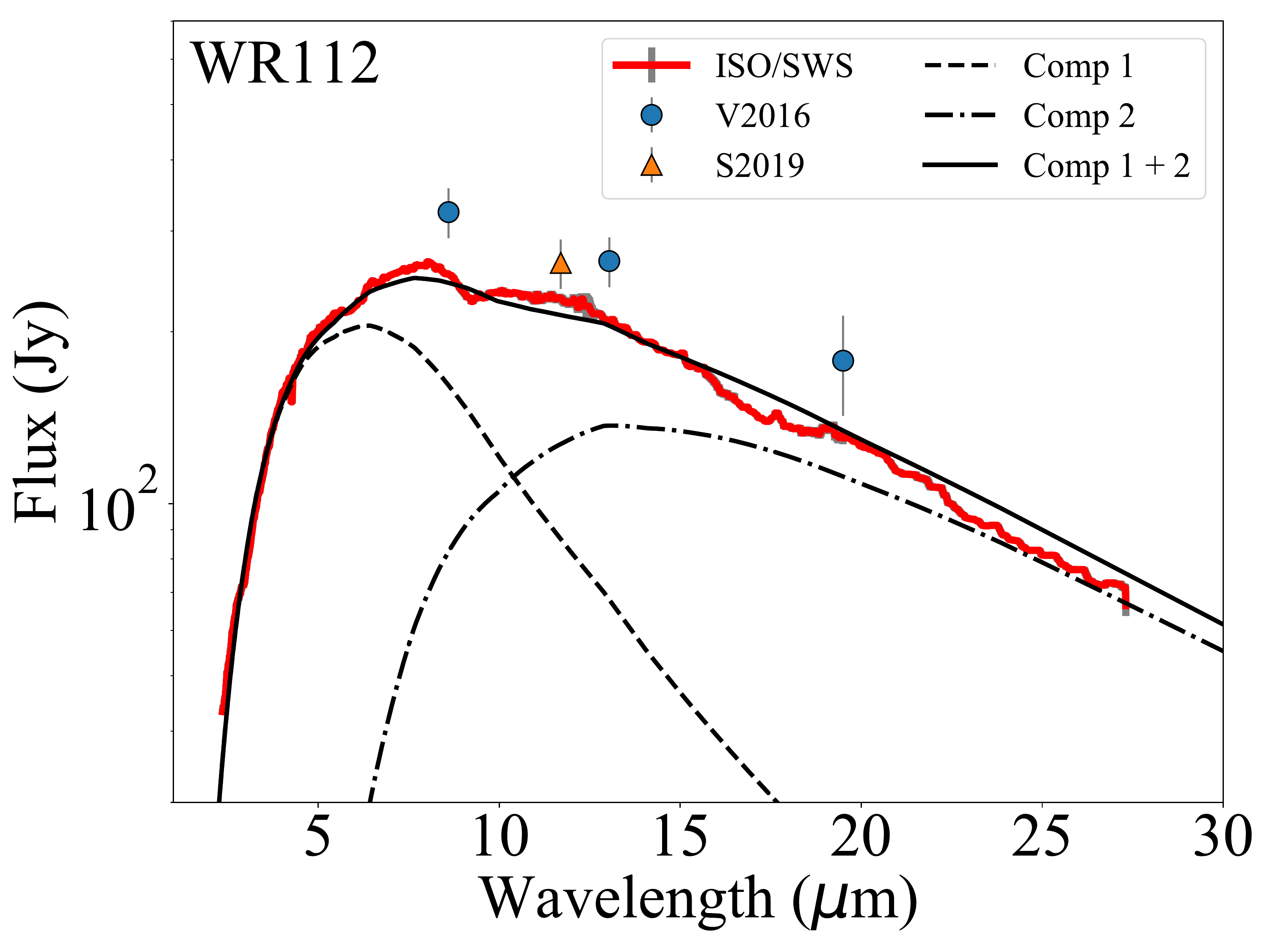}}
    \caption{Extinction-corrected ISO/SWS spectrum of WR~112 taken in 1996 Feb (red line) overlaid with the dereddened N-band photometry from V2016 (\citealt{Lau2017}; blue circles) and S2019 (orange triangle) and the best-fit 2-component dust emission mode. The emission from components 1 and 2 are shown as the dashed and dot-dashed lines, respectively, and the total combined model emission shown as the solid black line.}
    \label{fig:WR112SED}
\end{figure}

In this section, we revisit the dust production properties of WR~112 based on our proper motion analysis, \rev{revised distance estimate}, and \rev{new} geometric model.
In order to determine the dust properties, we fit a 2-component optically thin dust emission model to an extinction corrected ISO/SWS 2.2 - 27.5 $\mu$m spectrum of WR~112 (Fig.~\ref{fig:WR112SED}). Reddening by interstellar extinction is corrected using the ISM extinction law derived by \citet{Chiar2006} normalized to the visual extinction measured towards WR~112 of $A_V=11.03$\footnote{This is consistent with the extinction derived by \citet{vdh2001} where $A_v\approx 1.1 A_V$}. Initially, a single component dust model was attempted but resulted in an unsatisfactory fit due to the broad shape of the SWS spectrum. We note that the measured flux \rev{\revv{density}} from the SWS observations, which were taken in 1996 Feb, are consistent within $\sim20\%$ of the ground-based V2016 \citep{Lau2017} and S2019 mid-IR photometry. This supports the interpretation of WR~112 as a continuous dust producer with minimal variability in dust production.

We produce dust emission models using \dustem, a numerical tool that computes the dust emission in the optically thin limit heated by an input radiation field with no radiative transfer \citep{Compiegne2011}. Our 2-component dust model technique is identical to the WC dustar spectral energy distribution (SED) analysis by \citet{Lau2020}. The two components are treated as circular dust rings with a different radii heated radiatively by a central heating source, where the inner dust ring ($r_1$) attenuates the radiation heating the outer dust ring ($r_2$). Since the circumstellar material around WC \rev{dustars} is believed to be composed of amorphous carbon dust \citep{Cherchneff2000}, amorphous carbon grains with optical properties described by \citet{Zubko1996, Compiegne2011} are adopted for the \dustem~model. We use a grain size distribution with radii ranging from $a = 0.1 - 1.0$ $\mu$m and a number density distribution proportional to $n(a)\propto a^{-3}$. This grain size distribution is consistent with the $\sim0.5 - 1.0$ $\mu$m grain radii estimated in the WR~112 nebula by previous studies \citep{Chiar2001,Marchenko2002}.

\begin{deluxetable}{lr}
\tablecaption{WR~112 Distance and Dust Model Results}
\tablewidth{0.9\columnwidth}
\tablehead{Parameter & Value  }
\startdata
\rev{$d$} & \rev{$3.39^{+0.89}_{-0.84}$ kpc} \\
$r_1$ & \rev{$140^{+40}_{-50}$} AU\\
$r_2$ & \rev{$910^{+240}_{-350}$} AU\\ 
$T_\mathrm{d1}$ & \rev{$830^{+150}_{-70}$} K\\
$T_\mathrm{d2}$ & \rev{$300^{+100}_{-40}$} K\\
$L_\mathrm{IR}$ & \revv{$(6.6^{+0.3}_{-0.1})^{+3.9}_{-2.9}\times10^4$ L$_\odot$}\\
$M_\mathrm{d1}$ & \revv{$(1.4^{+1.0}_{-1.0})^{+0.9}_{-0.6}\times10^{-6}$ M$_\odot$}\\
$M_\mathrm{d2}$ & \revv{$(8.2^{+4.7}_{-4.9})^{+4.9}_{-3.6}\times10^{-5}$ M$_\odot$}\\
$\dot{M}_\mathrm{d}$ & \revv{$(2.7^{+1.0}_{-1.3})^{+1.6}_{-1.2}\times10^{-6}$ M$_\odot$ yr$^{-1}$}\\
$\chi_C$ ($4\pi$ steradians) & \revv{$(6.5^{+2.3}_{-3.0})^{+0.8}_{-0.9}  \,\%$}\\
$\chi_C$ ($\pm\theta_s$ Equatorial Band)& \rev{$\sim8 \,\%$}\\ 
\enddata
\tablecomments{Summary of the \rev{distance $d$ and} dust model results of WR~112, where the following values are \rev{adopted}: \rev{$v_\mathrm{exp}=1230$ km s$^{-1}$}, \rev{$\dot{M} = 1.1 \times10^{-4}$ M$_\odot$ yr$^{-1}$} \citep{Monnier2002}. The radius, temperature, and mass of the model components 1 and 2 are given $r_{1/2}$, $T_{d1/d2}$, and $M_{d1/d2}$, respectively. $L_\mathrm{IR}$ is the total IR luminosity of both dust components. The dust production rate $\dot{M}_\mathrm{d}$ and the dust condensation fraction of available carbon the WC wind $\chi_C$ are derived using the expansion derived by $v_\mathrm{exp}^\mathrm{PM}$, $M_\mathrm{d1}$, and $r_1$. \rev{The first and second set of uncertainties provided for $L_\mathrm{IR}$, $M_\mathrm{d1}$, $M_\mathrm{d2}$, $\dot{M}_\mathrm{d}$, and $\chi_C$ correspond to the $1\sigma$ uncertainties from the SED model fit and the WR~112 distance estimate, respectively. The distance to WR~112 does not impact the values derived for $r_{1/2}$, $T_{d1/d2}$ and thus only the SED model-fit uncertainties are shown for these quantities.}}
\label{tab:WR112Results}
\end{deluxetable}

The luminosity of the central system is not well-characterized given the high interstellar extinction. We therefore adopt a total system luminosity of \rev{$L_*=4.0\times10^5$ L$_\odot$} in agreement with the mean \rev{WC8} stellar luminosity determined by \citet{Sander2019} and assume that the WC star dominates the radiative heating. A Potsdam Wolf-Rayet Star (PoWR) model atmosphere \citep{Grafener2002,Sander2012,Sander2019} with an effective temperature of \rev{$T_*=50000$ K} and ``transformed" radius of \rev{7.9 R$_\odot$} was adopted for the radiation field of the WR~112 heating source, which is consistent with the values derived from the spectral pseudo-fit to the dusty \rev{WC8} system \rev{WR~53} by \rev{\citet{Sander2019}}.

Our best-fit 2-component model provides dust component distances of \rev{$r_1 = 140^{+40}_{-50}$} AU and \rev{$r_2 = 910^{+240}_{-350}$} AU corresponding to components 1 and 2, respectively. The dust temperature of the two components are \rev{$T_{d1}=830^{+150}_{-70}$} K and \rev{$T_{d2}=300^{+100}_{-40}$} K, and the total integrated IR luminosity of both components is \revv{$L_\mathrm{IR}=6.6^{+0.3}_{-0.1}\times10^4$} L$_\odot$. High spatial resolution 2.2 $\mu$m observations of WR~112 by \citet{Ragland1999} revealed an inner dust radius of 31 mas, which corresponds to a size of \rev{$\sim110$} AU at a distance of \rev{$d = 3.39$} kpc. The resolved inner dust radius and our model-derived component 1 radius are therefore in close agreement. We also note that component 1 dominates the emission at 2.2 $\mu$m (Fig.~\ref{fig:WR112SED}) and is thus consistent with the emitting dust component resolved by \citet{Ragland1999}.

From the 2-component model, we derive dust masses of \revv{$M_{d1}=1.4^{+1.0}_{-1.0}\times10^{-6}$} M$_\odot$ and \revv{$M_{d2}=8.2^{+4.7}_{-4.9}\times10^{-5}$} M$_\odot$ for components 1 and 2, respectively. Using the dust distances and masses, the dust production rate can be approximated by 

\begin{equation}
       \dot{M}_d \sim \frac{M_d \,v_\mathrm{exp}}{r}.
       \label{eq:continuous}
\end{equation}

\noindent
We adopt a conservative estimate of the dust production rate based only on the dust mass and the distance to component 1 since cooler and more extended dust may contribute to the dust mass determined for component 2. \rev{Assuming the dust expansion velocity, $v_\mathrm{exp}$, is equal to the terminal wind velocity of the WC star $v_\infty=1230$ km s $^{-1}$ (See Sec.~\ref{Sec:dist}),} the dust production rate based on component 1 is \revv{$\dot{M}_d=2.7^{+1.0}_{-1.3}\times10^{-6}$} M$_\odot$ yr$^{-1}$. \rev{The WR~112 dust model results are summarized in Tab.~\ref{tab:WR112Results}. We note that only the SED model-fitting uncertainties are provided in the text of this section, whereas Tab.~\ref{tab:WR112Results} lists both the model-fitting uncertainties and the uncertainties due to our distance estimate.}

\begin{figure*}[t!]
   \centerline{\includegraphics[width=1.0\linewidth]{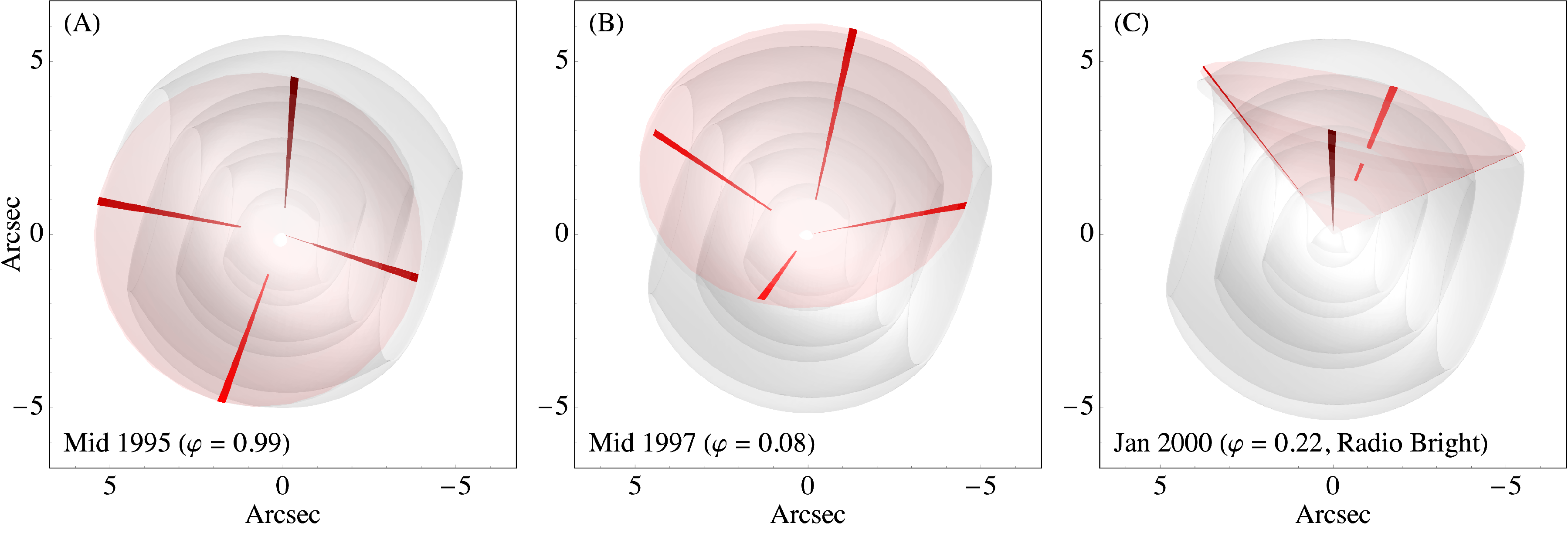}}
    \caption{3D surface model of WR~112 (grey) overlaid with the projection of the colliding-wind interface of the central binary (red) \rev{at our predicted viewing angle} in (A) mid 1995, (B) mid 1997, and (C) \rev{Jan 2000}. \rev{In each panel, red bands are overlaid on the surface of the shock cone to help illustrate the orientation and geometry}. \rev{WR~112 was observed in a radio-bright state between 1999 Sept and 2000 Feb} \citep{Monnier2002}.}
    \label{fig:WR112Schem}
\end{figure*}

The total mass-loss rate of the WC wind in WR~112 derived from its thermal radio emission by \citet{Monnier2002} is \rev{$1.1\times10^{-4}$} M$_\odot$ yr$^{-1}$, which has been re-scaled to \rev{our revised $d = 3.39$ kpc} distance. \rev{By assuming a carbon mass fraction of $40\%$ in WC winds \citep{Sander2012,Sander2019}, we estimate a carbon dust mass condensation fraction in WR~112 of \revv{$\chi_C=6.5^{+2.3}_{-3.0}\%$}}. Note that this is the condensation fraction for the total mass-loss rate of WR~112 over $4\pi$ steradians. However, the region over which the WC wind can contribute to dust formation is constrained to the equatorial angular band within $\pm\theta_s$ of the orbital plane \citep{Tuthill2008}. For $\theta_s=55^\circ$, this equatorial angular band is $82\%$ of the $4\pi$ steradians subtended by the WC wind in WR~112. Therefore, \rev{$\sim8\%$} of the carbon by mass from the WC wind in this equatorial band condenses to dust. These results provide strong evidence that dust can form efficiently in the colliding winds of WC binaries.

\section{Discussion}
\label{sec:4}

\subsection{Reconciling WR~112 Non-thermal Radio Variability}
With the revised geometric spiral model of WR~112, we can resolve lingering uncertainties on the nature of WR~112 posed by previous radio observations (e.g.~\citealt{Monnier2002}). WR~112 exhibited highly variable non-thermal radio emission, which is difficult to reconcile with a face-on geometry \citep{Monnier2002,Monnier2007}. Non-thermal radio emission arises from particle acceleration in the shock collision zone \rev{between the winds of the WC and OB-star}; however, the non-thermal radio emission can be \rev{absorbed by the dense, ionized} stellar winds of the WC and OB star \citep{Eichler1993}. Variability in the non-thermal radio emission can therefore arise from the orbital motion of a near edge-on central binary due to changes in the optical depth along the line of sight to the wind collision zone.

 \rev{The variable and highly absorbed non-thermal radio emission from WR~112 was revealed by \citet{Chapman1999} who measured a 2.38 GHz flux \rev{\revv{density}} of $F_\mathrm{2.38GHz}=3.8$ mJy in 1997, which was a factor of 10 greater than the expected 2.38 GHz flux \rev{\revv{density}} based on a purely thermal emission model and a 8.64 GHz flux \rev{\revv{density}} measurement of $F_\mathrm{8.64GHz}=0.68$ mJy in 1995 by \citet{Leitherer1997}. \citet{Monnier2002} showed that WR112 was in a radio bright state between 1999 Sept and 2000 Feb, where the 1.425 GHz flux \rev{\revv{density}} was measured to be $\sim2.5$ mJy while the 1.38 GHz flux \rev{\revv{density}} measurement by \citet{Chapman1999} in 1997 provided an upper limit of $<1.1$ mJy. Notably, radio observations of WR~112 by \citet{Monnier2002} implied a shallower slope between the 2.38 and 1.38 GHz flux \rev{\revv{density}} measurements by \citet{Chapman1999}, which suggests a lower free-free opacity in the \citet{Monnier2002} observations.}

In Fig.~\ref{fig:WR112Schem}, we show three \rev{orbital phases} of the 3D spiral \rev{model} and the projection of the colliding-wind interface corresponding to different dates between 1995 and 2000. The projected shock cone shown in Fig.~\ref{fig:WR112Schem} is the 3D surface of the shock interface shown in Fig.~\ref{fig:CWSchem} and is analogous to the conic surface \~C in \citet{Eichler1993}. Note that the shock cone surface in Fig.~\ref{fig:WR112Schem} does not incorporate the influence of the binary orbital motion since this effect on the skew angle of the shock cone is negligible (See Sec.~\ref{sec:CW}). 

Based on our geometric model, our line of sight towards WR~112 was aligned with the interior of the shock cone between 1995 and 1997\rev{, which correspond to orbital phases $\varphi \sim 0$ and $0.1$, respectively} (Fig.~\ref{fig:WR112Schem}A \& B). Non-thermal radio variability is expected in this orientation where optically thick winds from the OB companion can obscure the shock collision zone \citep{Eichler1993, Dougherty2003}.
During the radio bright state between late 1999 and early 2000, our model \rev{at 2000 Jan is consistent with an orbital phase of $\varphi = 0.22$, which is near quadrature. Non-thermal radio emission models of a WR+OB colliding wind system by \citet{Dougherty2003} demonstrate that attenuation from free-free opacity is weakest near quadrature. At this phase where the WC and OB star are aligned parallel to the plane of the sky, our line of sight towards the non-thermal emission region at the apex of the shock interface is less obscured by the stellar winds from both the WR star and its companion.}
\rev{The revised near edge-on geometry} of the WR~112 system is therefore consistent with the observed radio variability. 

\rev{The subsequent quadrature where WR~112 should have exhibited another radio bright state would have occurred in 2010, which is half an orbital phase ( $\sim10$ yr) after the observed radio bright state in late 1999 - early 2000. Interestingly, the radio light curve of WR~112 presented by \citet{Yam2015} reveals an 8.4 GHz emission peak from $\sim2010 - 2011$, which is consistent with our model prediction\revv{; however, the flux density measured by \citet{Yam2015} during this peak is a factor of $\sim3$ less than the 8.46 GHz emission ($F_\mathrm{8.46GHz}=4.07$) reported by \citet{Monnier2002} in 2000 Feb}. WR~112's next radio bright state should occur during the following quadrature expected in $\sim2020-2021$.}

\subsection{WR~112 and WC Dustar Diversity}
WR~112 exhibits one of the highest dust production rates of the known WC dustars, which range from $\sim10^{-10} - 10^{-6}$ M$_\odot$ yr$^{-1}$ \citep{Lau2020}. Despite their contrasting orbital and colliding-wind shock properties, WR~112 exhibits a dust production rate comparable to one of the heaviest known dust-makers WR~104 ($\sim4\times10^{-6}$ M$_\odot$ yr$^{-1}$; \citealt{Lau2020}).
A comparison between WR~112 and WR~104 therefore highlights the range of the WC binary orbital parameters that exhibit high dust production rates ($\sim10^{-6}$ M$_\odot$ yr$^{-1}$). To provide a comparative benchmark, this dust production rate exceeds the total measured dust input from the AGB stars in the Small Magellanic Cloud ($\sim9\times10^{-7}$ M$_\odot$ yr$^{-1}$; \citealt{Boyer2012}). Both WR~112 and WR~104 host \rev{late-type WC}+OB binaries with near circular orbits. However, WR~104 is a near face-on system with an orbital separation of $\sim2$ AU and a relatively narrow colliding-wind half-opening angle $\theta_s\approx20^\circ$ \citep{Tuthill2008} while WR~112 is a near edge-on system with an orbital separation of $\sim20$ AU and a wide colliding wind half-opening $\theta_s=55^\circ$. 

Since dust formation is linked to the wind interaction with the binary companion, it is interesting compare the inferred properties of the OB-companions in WR~112 and WR~104. \citet{Harries2004} propose that WR~104 hosts an OB-companion that exhibits a mass-loss rate of $6\times10^{-8}$ M$_\odot$ yr$^{-1}$ with an assumed wind velocity of $2000$ km s$^{-1}$. Based on the mass-loss rate and wind velocity of the WC star in WR~112 and the wind-momentum ratio $\eta\approx0.13$ derived from its observed opening angle (Eq.~\ref{eq:eta}), we estimate a mass-loss rate of \rev{$\sim8\times10^{-6}$ M$_\odot$ yr$^{-1}$} for the OB-companion with an assumed wind velocity of $2000$ km s$^{-1}$. This high mass-loss rate would be consistent with an early O-type supergiant (e.g.~\citealt{Muijres2012}). \rev{Companions with such} high mass-loss rates may be necessary to enable dust formation in systems with large orbital separations\rev{/periods} such as WR~112, WR~48a (P$_\mathrm{orb}=32.5$ yr; \citealt{Williams2012}), and Apep (P$_\mathrm{orb}\sim100$ yr; \citealt{Han2020}). Both WR~48a and Apep indeed host companions that exhibit high mass-loss rates \citep{Zhekov2014,Callingham2020}. \rev{Interestingly, WR~112, WR~48a, and Apep exhibit WR emission lines consistent with a WC8 star.} 

The range of orbital separations \rev{exhibited by} WC dustars \rev{and possible trends with their spectral sub-types} highlight the importance of investigating the relation between \rev{their} dust formation efficiency and stellar and orbital properties. 
\rev{The longer orbital period of WR~112 relative to other heavy WC dust makers with $\sim$yr periods like WR~104 also highlights the diversity of WC systems that exhibit high dust formation rates and bolsters their likely role as significant sources of dust in the ISM \citep{Lau2020}.}

\section{Conclusions}

We have presented a multi-epoch morphological analysis of the complex geometry of the dusty circumstellar environment formed by the colliding-wind binary WR~112. \rev{Our} analysis utilized high resolution N-band imaging observations of WR~112 taken over almost a 20-yr baseline with Gemini North/OSCIR, Gemini South/T-ReCS, Keck I/LWS, VLT/VISIR, and Subaru/COMICS (Fig.~\ref{fig:WR112Comics} \&~\ref{fig:WR112Im}). WR~112's changing dust morphology is consistent with a nearly edge-on 3D conical spiral with a wide opening angle that appears to rotate due to persistent dust formation from a central colliding-wind WC binary with a $\sim20$-yr orbital period. The motion of the dust, however, is not along the spiral but is radial since dust forms and propagates in the changing direction of the shock cone between the WC star and its companion. 

The observed dust emission morphology and the dust column density model image derived from the revised geometry show a close match (Fig.~\ref{fig:WR112Mod}), which supports our revised interpretation of the WR~112 nebula geometry. Our geometric model and the inferred $\sim20$-yr orbital period provides us with insight on the orbital configuration of the central binary and the shock cone produced by the wind collision.

We revisited the \rev{distance estimate and} dust production properties of WR~112 with this revised geometric model and estimate high dust production rates and dust condensation efficiency. With a dust production rate of \rev{$\sim3\times10^{-6}$ M$_\odot$ yr$^{-1}$}, WR~112 is among the most prolific dust makers of all known WC dustars and highlights their impact as significant dust producers in the ISM.

The nearly edge-on interpretation of WR~112 reconciles inconsistencies highlighted by previous observations of variable non-thermal radio emission from the wind collision zone. Continued radio observations and future work on the correlation between the mid-IR dust morphology and radio light curve would therefore provide valuable insight on the wind collision region and properties of the stellar winds. \rev{Additionally, future observations utilizing very long baseline interferometry may be able to reveal the morphology of non-thermal emission (e.g.~\citealt{Dougherty2005,SB2019}).}

Given to its IR brightness, resolvable extended emission, and high dust formation rate, WR~112 presents a unique laboratory to investigate dust formation and wind-collision dynamics in colliding wind binaries. State-of-the-art hydrodynamic simulations and 3D radiative transfer models (e.g.~\citealt{Lamberts2012,Hendrix2016,Soulain2018,Calderon2020}) of WR~112 would deepen our understanding of such phenomena. WR~112's relatively extended emission compared to the nebulae around shorter period dusty WC systems also provides a valuable reference for interpreting the morphologies of other dust-forming colliding-wind binaries. Lastly, we emphasize the importance of multi-epoch high spatial resolution mid-IR observations for pursuing these studies. Mid-IR instrumentation on future observatories such as the Tokyo Atacama Observatory \citep{Miyata2010}, 30-m class telescopes, and the \textit{James Webb Space Telescope} will be crucial for unraveling the nature of dust formation in these efficient dust factories.

\acknowledgments
\textit{Acknowledgements.}
RML thanks D.~Calder\'{o}n,  A.~Lamberts, S.~V.~Marchenko, J.~D.~Monnier, \rev{T.~Onaka}, G.~Rate, C.~M.~P.~Russell, and R.~Sch\"{o}del for the valuable discussion on colliding-winds in WC binaries and observations of WR~112. 
\rev{We thank the anonymous referee for the insightful comments and suggestions that have improved the clarity and quality of this work.}
RML also thanks T.~Fujiyoshi and the Subaru Observatory staff for supporting our Subaru/COMICS observations of WR~112.  
RML acknowledges the Japan Aerospace Exploration Agency's International Top Young Fellowship.
AFJM is grateful for financial aid from NSERC (Canada).
Based in part on observations collected at the European Organisation for Astronomical Research in the Southern Hemisphere under ESO programme 097.D-0707(A).
Some of the data presented herein were obtained at the W. M. Keck Observatory, which is operated as a scientific partnership among the California Institute of Technology, the University of California and the National Aeronautics and Space Administration. The Observatory was made possible by the generous financial support of the W. M. Keck Foundation.
The authors wish to recognize and acknowledge the very significant cultural role and reverence that the summit of Maunakea has always had within the indigenous Hawaiian community.  We are most fortunate to have the opportunity to conduct observations from this mountain.
Based in part on data collected at Subaru Telescope, which is operated by the National Astronomical Observatory of Japan.
Based in part on observations obtained at the international Gemini Observatory, a program of NOIRLab, which is managed by the Association of Universities for Research in Astronomy (AURA) under a cooperative agreement with the National Science Foundation on behalf of the Gemini Observatory partnership: the National Science Foundation (United States), National Research Council (Canada), Agencia Nacional de Investigaci\'{o}n y Desarrollo (Chile), Ministerio de Ciencia, Tecnolog\'{i}a e Innovaci\'{o}n (Argentina), Minist\'{e}rio da Ci\^{e}ncia, Tecnologia, Inova\c{c}\~{o}es e Comunica\c{c}\~{o}es (Brazil), and Korea Astronomy and Space Science Institute (Republic of Korea).
Based in part on observations with ISO, an ESA project with instruments funded by ESA Member States (especially the PI countries: France, Germany, the Netherlands and the United Kingdom) and with the participation of ISAS and NASA.
Based in part on observations in the service observing programme of the WHT, operated on the island of La Palma by the Isaac Newton Group of Telescopes in the Spanish Observatorio del Roque de los Muchachos of the Instituto de Astrofísica de Canarias. 
This research made use of Astropy,\footnote{http://www.astropy.org} a community-developed core Python package for Astronomy \citep{Astropy2013, Astropy2018}.

%

\vspace{5mm}
\facilities{VLT(VISIR), Subaru(COMICS), Gemini South(T-ReCS), Keck I(LWS), Gemini North(OSCIR), \rev{ISO(SWS)}, WHT(ISIS)}


\begin{thebibliography}{}


\bibitem[Astropy Collaboration et al.(2013)]{Astropy2013} Astropy Collaboration, Robitaille, T.~P., Tollerud, E.~J., et al.\ 2013, \aap, 558, A33

\bibitem[Astropy Collaboration et al.(2018)]{Astropy2018} Astropy Collaboration, Price-Whelan, A.~M., Sip{\H{o}}cz, B.~M., et al.\ 2018, \aj, 156, 123


\bibitem[Boyer et al.(2012)]{Boyer2012} Boyer, M.~L., Srinivasan, S., Riebel, D., et al.\ 2012, \apj, 748, 40


\bibitem[Calder{\'o}n et al.(2020)]{Calderon2020} Calder{\'o}n, D., Cuadra, J., Schartmann, M., et al.\ 2020, \mnras, 493, 447


\bibitem[Callingham et al.(2019)]{Callingham2019} Callingham, J.~R., Tuthill, P.~G., Pope, B.~J.~S., et al.\ 2019, Nature Astronomy, 3, 82

\bibitem[Callingham et al.(2020)]{Callingham2020} Callingham, J.~R., Crowther, P.~A., Williams, P.~M., et al.\ 2020, \mnras, 495, 3323


\bibitem[Cant\'{o} et al.(1996)]{Canto1996} Cant\'{o}, J., Raga, A.~C., \& Wilkin, F.~P.\ 1996, \apj, 469, 729

\bibitem[Chiar \& Tielens(2001)]{Chiar2001} Chiar, J.~E., \& Tielens, A.~G.~G.~M.\ 2001, \apjl, 550, L207

\bibitem[Chiar \& Tielens(2006)]{Chiar2006} Chiar, J.~E., \& Tielens, A.~G.~G.~M.\ 2006, \apj, 637, 774

\bibitem[Chapman et al.(1999)]{Chapman1999} Chapman, J.~M., Leitherer, C., Koribalski, B., et al.\ 1999, \apj, 518, 890

\bibitem[Cherchneff et al.(2000)]{Cherchneff2000} Cherchneff, I., Le Teuff, Y.~H., Williams, P.~M., et al.\ 2000, \aap, 357, 572

\bibitem[Cohen \& Kuhi(1976)]{Cohen1976} Cohen, M., \& Kuhi, L.~V.\ 1976, \pasp, 88, 535


\bibitem[Cohen et al.(1999)]{Cohen1999} Cohen, M., Walker, R.~G., Carter, B., et al.\ 1999, \aj, 117, 1864

\bibitem[Compi{\`e}gne et al.(2011)]{Compiegne2011} Compi{\`e}gne, M., Verstraete, L., Jones, A., et al.\ 2011, \aap, 525, A103

\bibitem[Crowther et al.(1998)]{Crowther1998} Crowther, P.~A., De Marco, O., \& Barlow, M.~J.\ 1998, \mnras, 296, 367

\bibitem[Crowther(2007)]{Crowther2007} Crowther, P.~A.\ 2007, \araa, 45, 177

\bibitem[Currie et al.(2014)]{Starlink} Currie, M.~J., Berry, D.~S., Jenness, T., et al.\ 2014, Astronomical Data Analysis Software and Systems XXIII, 391


\bibitem[Cutri et al.(2003)]{Cutri2003} Cutri, R.~M., Skrutskie, M.~F., van Dyk, S., et al.\ 2003, VizieR Online Data Catalog, 2246,

\bibitem[De Buizer \& Fisher(2005)]{trecs} De Buizer, J., \& Fisher, R.\ 2005, High Resolution Infrared Spectroscopy in Astronomy, 84

\bibitem[de Graauw et al.(1996)]{deGraauw1996} de Graauw, T., Haser, L.~N., Beintema, D.~A., et al.\ 1996, \aap, 315, L49 

\bibitem[Dougherty et al.(2003)]{Dougherty2003} Dougherty, S.~M., Pittard, J.~M., Kasian, L., et al.\ 2003, \aap, 409, 217

\bibitem[Dougherty et al.(2005)]{Dougherty2005} Dougherty, S.~M., Beasley, A.~J., Claussen, M.~J., et al.\ 2005, \apj, 623, 447

\bibitem[Eichler \& Usov(1993)]{Eichler1993} Eichler, D., \& Usov, V.\ 1993, \apj, 402, 271


\bibitem[Gehrz \& Hackwell(1974)]{Gehrz1974} Gehrz, R.~D., \& Hackwell, J.~A.\ 1974, \apj, 194, 619


\bibitem[Gr{\"a}fener et al.(2002)]{Grafener2002} Gr{\"a}fener, G., Koesterke, L., \& Hamann, W.-R.\ 2002, \aap, 387, 244 

\bibitem[Han et al.(2020)]{Han2020} Han, Y., et al.\ 2020, \mnras, Submitted

\bibitem[Harries et al.(2004)]{Harries2004} Harries, T.~J., Monnier, J.~D., Symington, N.~H., et al.\ 2004, \mnras, 350, 565


\bibitem[Hendrix et al.(2016)]{Hendrix2016} Hendrix, T., Keppens, R., van Marle, A.~J., et al.\ 2016, \mnras, 460, 3975

\bibitem[Jones \& Puetter(1993)]{LWS} Jones, B., \& Puetter, R.~C.\ 1993, \procspie, 610

\bibitem[Kataza et al.(2000)]{Kataza2000} Kataza, H., Okamoto, Y., Takubo, S., et al.\ 2000, \procspie, 1144

\bibitem[Kessler et al.(1996)]{Kessler1996} Kessler, M.~F., Steinz, J.~A., Anderegg, M.~E., et al.\ 1996, \aap, 500, 493

\bibitem[Lagage et al.(2004)]{Lagage2004} Lagage, P.~O., Pel, J.~W., Authier, M., et al.\ 2004, The Messenger, 117, 12

\bibitem[Lamberts et al.(2012)]{Lamberts2012} Lamberts, A., Dubus, G., Lesur, G., et al.\ 2012, \aap, 546, A60

\bibitem[Lau et al.(2017)]{Lau2017} Lau, R.~M., Hankins, M.~J., Sch{\"o}del, R., et al.\ 2017, \apjl, 835, L31


\bibitem[Lau et al.(2020)]{Lau2020} Lau, R.~M., Eldridge, J.~J., Hankins, M.~J., et al.\ 2020, \apj, 898, 74

\bibitem[Leitherer et al.(1997)]{Leitherer1997} Leitherer, C., Chapman, J.~M., \& Koribalski, B.\ 1997, \apj, 481, 898


\bibitem[Marchenko et al.(2002)]{Marchenko2002} Marchenko, S.~V., Moffat, A.~F.~J., Vacca, W.~D., et al.\ 2002, \apjl, 565, L59

\bibitem[Marchenko \& Moffat(2007)]{Marchenko2007} Marchenko, S.~V., \& Moffat, A.~F.~J.\ 2007, Massive Stars in Interactive Binaries, 367, 213 

\bibitem[Marchenko \& Moffat(2017)]{Marchenko2017} Marchenko, S.~V., \& Moffat, A.~F.~J.\ 2017, \mnras, 468, 2416 

\bibitem[Massey \& Conti(1983)]{Massey1983} Massey, P., \& Conti, P.~S.\ 1983, \pasp, 95, 440

\bibitem[Miyata et al.(2010)]{Miyata2010} Miyata, T., Sako, S., Nakamura, T., et al.\ 2010, \procspie, 77353P

\bibitem[Monnier et al.(1999)]{Monnier1999} Monnier, J.~D., Tuthill, P.~G., \& Danchi, W.~C.\ 1999, \apjl, 525, L97

\bibitem[Monnier et al.(2002)]{Monnier2002} Monnier, J.~D., Greenhill, L.~J., Tuthill, P.~G., et al.\ 2002, \apj, 566, 399

\bibitem[Monnier et al.(2007)]{Monnier2007} Monnier, J.~D., Tuthill, P.~G., Danchi, W.~C., et al.\ 2007, \apj, 655, 1033

\bibitem[Muijres et al.(2012)]{Muijres2012} Muijres, L.~E., Vink, J.~S., de Koter, A., et al.\ 2012, \aap, 537, A37


\bibitem[Murakawa et al.(2004)]{Murakawa2004} Murakawa, K., Suto, H., Tamura, M., et al.\ 2004, \pasj, 56, 509

\bibitem[Nugis et al.(1998)]{Nugis1998} Nugis, T., Crowther, P.~A., \& Willis, A.~J.\ 1998, \aap, 333, 956

\bibitem[Okamoto et al.(2003)]{Okamoto2003} Okamoto, Y.~K., Kataza, H., Yamashita, T., et al.\ 2003, \procspie, 169

\bibitem[Parkin \& Pittard(2008)]{Parkin2008} Parkin, E.~R., \& Pittard, J.~M.\ 2008, \mnras, 388, 1047


\bibitem[Ragland \& Richichi(1999)]{Ragland1999} Ragland, S., \& Richichi, A.\ 1999, \mnras, 302, L13

\bibitem[Rate \& Crowther(2020)]{RC2020} Rate, G., \& Crowther, P.~A.\ 2020, \mnras, 493, 1512

\bibitem[Sanchez-Bermudez et al.(2019)]{SB2019} Sanchez-Bermudez, J., Alberdi, A., Sch{\"o}del, R., et al.\ 2019, \aap, 624, A55

\bibitem[Sander et al.(2012)]{Sander2012} Sander, A., Hamann, W.-R., \& Todt, H.\ 2012, \aap, 540, A144

\bibitem[Sander et al.(2019)]{Sander2019} Sander, A.~A.~C., Hamann, W.-R., Todt, H., et al.\ 2019, \aap, 621, A92 

\bibitem[Shortridge(1993)]{Figaro} Shortridge, K.\ 1993, Astronomical Data Analysis Software and Systems II, 219

\bibitem[Sloan et al.(2003)]{Sloan2003} Sloan, G.~C., Kraemer, K.~E., Price, S.~D., \& Shipman, R.~F.\ 2003, \apjs, 147, 379 

\bibitem[Soulain et al.(2018)]{Soulain2018} Soulain, A., Millour, F., Lopez, B., et al.\ 2018, \aap, 618, A108 


\bibitem[Tuthill et al.(1999)]{Tuthill1999} Tuthill, P.~G., Monnier, J.~D., \& Danchi, W.~C.\ 1999, \nat, 398, 487 

\bibitem[Tuthill et al.(2008)]{Tuthill2008} Tuthill, P.~G., Monnier, J.~D., Lawrance, N., et al.\ 2008, \apj, 675, 698 

\bibitem[Wallace et al.(2002)]{Wallace2002} Wallace, D.~J., Moffat, A.~F.~J., \& Shara, M.~M.\ 2002, Interacting Winds from Massive Stars, 407

\bibitem[Williams et al.(1987)]{Williams1987} Williams, P.~M., van der Hucht, K.~A., \& Th\'{e}, P.~S.\ 1987, \aap, 182, 91


\bibitem[Williams et al.(1990)]{Williams1990} Williams, P.~M., van der Hucht, K.~A., Pollock, A.~M.~T., et al.\ 1990, \mnras, 243, 662

\bibitem[Williams et al.(2012)]{Williams2012} Williams, P.~M., van der Hucht, K.~A., van Wyk, F., et al.\ 2012, \mnras, 420, 2526


\bibitem[Williams \& van der Hucht(2015)]{Williams2015} Williams, P.~M., \& van der Hucht, K.~A.\ 2015, Wolf-rayet Stars, 275


\bibitem[Usov(1991)]{Usov1991} Usov, V.~V.\ 1991, \mnras, 252, 49 


\bibitem[van der Hucht et al.(1996)]{vdh1996} van der Hucht, K.~A., Morris, P.~W., Williams, P.~M., et al.\ 1996, \aap, 315, L193 

\bibitem[van der Hucht(2001)]{vdh2001} van der Hucht, K.~A.\ 2001, \nar, 45, 135 

\bibitem[Yam et al.(2015)]{Yam2015} Yam, J.~O., Dzib, S.~A., Rodr{\'\i}guez, L.~F., et al.\ 2015, \rmxaa, 51, 35

\bibitem[Zhekov et al.(2014)]{Zhekov2014} Zhekov, S.~A., Gagn{\'e}, M., \& Skinner, S.~L.\ 2014, \apj, 785, 8


\bibitem[Zubko et al.(1996)]{Zubko1996} Zubko, V.~G., Mennella, V., Colangeli, L., et al.\ 1996, \mnras, 282, 1321
 
\bibitem[Zubko(1998)]{Zubko1998} Zubko, V.~G.\ 1998, \mnras, 295, 109


 
\end{thebibliography}
\end{document}